\documentclass[11pt]{article}
\usepackage{amssymb, graphicx, epsfig, amsmath, pifont, verbatim}
\graphicspath{{./}{/scratch/lenny/}{/scratch/weekly_reports/}{/scratch/}}
\makeatletter
\renewcommand*\@fnsymbol[1]{\the#1}
\makeatother
\newcommand{\ud}{\mathrm{d}}
\title{Initial Distribution Spread: A density forecasting approach}
\author{R. L. Machete\footnote{Corresponding author, email: {\tt r.l.machete@reading.ac.uk}, tel: +44(0)118 378 6378} $^{,a,b}$ and I. M. Moroz$^b$\\
{\footnotesize a. Department of Mathematics and Statistics, University of Reading, RG6 6AX, UK}\\{\footnotesize b. Mathematical Institute, 24-29 St Giles', Oxford, OX1 3LB, UK}}
\date{}
\begin{document}
\maketitle
\setcounter{footnote}{1}
\graphicspath{{./}{../graphs/}}
\begin{abstract}
Ensemble forecasting of nonlinear systems involves the use of a model to run forward a discrete ensemble (or set) of initial states. Data assimilation techniques tend to focus on estimating the true state of the system, even though model error limits the value of such efforts. This paper argues for choosing the initial ensemble in order to optimise forecasting performance rather than estimate the true state of the system. Density forecasting and choosing the initial ensemble are treated as one problem. Forecasting performance can be quantified by some scoring rule. In the case of the logarithmic scoring rule, theoretical arguments and empirical results are presented. It turns out that, if the underlying noise dominates model error, we can diagnose the noise spread.
\end{abstract}
{\renewcommand{\thefootnote}{}
\footnotetext{Abbreviations: European Centre for Medium-Range Weather Forecasts~(ECMWF)}}
{\bf Keywords}:
{data assimilation; density forecast; ensemble forecasting; uncertainty}
\section{Introduction} 
Given an {\it initial state} of some chaotic dynamical system $-$ examples of which include the population of an animal species in a game reserve, daily temperature for Botswana or day to day electricity demands for London $-$ we could perform a point forecast from the single state. Observational uncertainty and/or model error could limit the value of such a forecast. One can go over these hurdles by generating a discrete set of initial states in the neighbourhood of the current state and then forecasting from it.  The set of initial states is called an {\it initial ensemble}. Forecasting from an initial ensemble is called {\it ensemble forecasting}~\cite{leu-08}. The time ahead at which forecasts are made from any member of the initial ensemble is called the {\it lead time}. Ensemble forecasting is performed mainly to account for uncertainty in the initial conditions, although it can also be used to mitigate model error. A  lot of attention has been paid to generating initial ensembles~(e.g see~\cite{leu-08,kev-01}). 

Here, we present a novel approach to selecting the spread of the distribution from which an initial ensemble is drawn. The distribution from which an initial ensemble is drawn shall be called the {\it initial distribution}. Its covariance matrix will be taken to be spatially diagonal and uniform over all the initial conditions considered, in tune with common practise in data assimilation and ensemble forecasting~\cite{leu-08,lau-10,hol-86}. Note, however, that we do not assume that the initial distribution is the underlying noise distribution. We consider the choice of covariance structure to be a modelling decision to be made in a given situation. A diagonal covariance matrix is especially appealing when the number of state variables is large since it reduces computational complexity and costs. In weather forecasting, for example, one is confronted with a matrix of size $10^7\times 10^7$ and it is clearly undesirable for such a matrix to be full. There are, however, efforts to increase complexity to a tridiagonal matrix, but the effect of this on forecasting performance is yet unknown.

Ideally, the initial ensemble should be drawn from the underlying invariant measure, in which case we have a {\it perfect initial ensemble}. A perfect initial ensemble is especially useful in the scenario when our forecasting model is isomorphic to the model that generated the data, a scenario called the {\it perfect model scenario}~\cite{kev-01,kev-04}. When there is no isomorphism between the forecasting model and the model that generated the data, we are in the {\it imperfect model scenario}. A perfect model with a perfect initial ensemble would give us a {\it perfect forecast}~\cite{szf-99}. If either our model or initial ensemble is not perfect, then we have no reason to expect perfect forecasts.

In all realistic situations, we have neither a perfect model nor a perfect initial ensemble, yet we may be required to issue a meaningful forecast probability density function (pdf). Roulston and Smith~\cite{roul-03} proposed a methodology for making forecast distributions that are consistent with historical observations from ensembles. This is necessary because the forecast ensembles are not drawn from the underlying invariant measure due to either imperfect initial ensembles or model error. Their methodology was extended by Brocker and Smith~\cite{joc-07} to employ continuous density estimation techniques~\cite{par-62,sil-86} and blend the ensemble pdfs with the empirical distribution of historical data, which is referred to as the {\it climatology}. The resulting pdf is what will be taken as the forecast~pdf in this paper.

The quality of the forecast pdfs can be assessed  using the logarithmic scoring rule proposed by Good~\cite{good-52} and termed {\it Ignorance} by Roulston and Smith~\cite{roul-02}, borrowed from information theory~\cite{khi,sha}. Here, we discuss a way of choosing the initial distribution spread to enhance the quality of the forecast pdfs. The point is that if the spread is too small our forecasts may be over confident and if it is too large our forecasts may have low information content. Our goal is to choose an initial distribution spread that yields the most informative forecast pdfs and determine, for instance, if this varies with the lead time of interest. As is commonly done in data assimilation and ensemble forecasting~(e.g. see \cite{leu-08,joc-02}), we only consider Gaussian initial distributions. This distributional assumption can be relaxed to unimodal distributions. It is a reasonable assumption because it implies one is confident about some initial state. In traditional data assimilation and ensemble forecasting techniques, estimation of the initial distribution is divorced from forecasting: this is the main point of departure in our approach. We revisit this later in the discussion of the results in~\S~\ref{sec:disc}.

Our numerical forecasting experiments were performed on the Moore-Spiegel~\cite{mor-spi} system and an electronic circuit motivated by the Moore-Spiegel~system. Indeed electronic circuits have been studied to enhance our understanding of chaotic systems and Chua circuits~\cite{chu-92} are among famous examples. Recently, Gorlov and Strogonov~\cite{gor-07} applied ARIMA models to forecast the time to failure of Integrated Circuits. Hence, electronic circuits have not only been studied to enhance our understanding of chaotic systems and the forecasting of real systems, but also to understand the circuits themselves and to address practical design questions.

This paper is organised as follows: \S~\ref{sec:fore} introduces the technical framework for discussing probabilistic forecasting of deterministic systems. The theoretical and empirical scores for probabilistic forecasts are presented in~\S~\ref{sec:scor}. Density forecast estimation from ensemble forecasts is discussed in \S~\ref{sec:dfe}. Models used to produce forecasts are discussed in \S~\ref{sec:rbf}. Computations of the initial ensemble spread are discussed in \S~\ref{sec:init} for the perfect model scenario and imperfect model scenario. In the perfect model scenario, the Moore-Spiegel system~\cite{mor-spi} is considered. For the imperfect model scenario, the Moore-Spiegel system and an electronic circuit are modelled using radial basis function models. The circuit was constructed in a physics laboratory using state of the art equipment to mimic the Moore-Spiegel system. A theoretical argument in support of the numerical results is also presented. Implications and practical relevance of the results are discussed in~\S~\ref{sec:disc} and concluding remarks are given in \S~\ref{sec:con}.
\section{Forecasting}
\label{sec:fore}
Consider a deterministic dynamical system,
\begin{equation}
\dot{\boldsymbol{x}}=\boldsymbol{F}(\boldsymbol{x}(t)),
\label{s1:eq1}
\end{equation}
with the initial condition $\boldsymbol{x}(0)=\boldsymbol{x}_0$, where $\boldsymbol{x},\boldsymbol{F}\in\mathbb{R}^m$, $\boldsymbol{F}$ is a differentiable, nonlinear vector field and $t$ is time. By Picard's theorem~\cite{cod-55}, (\ref{s1:eq1}) will have a unique solution, say $\boldsymbol{x}(t)=\boldsymbol{\varphi}_t(\boldsymbol{x}_0)$. If $\nabla.\boldsymbol{F}<0$, this system might have an attractor~\cite{ott-94}, which, if it exists, we denote by $A$. We are interested in the case when the flow on this attractor is chaotic.
\subsection{Forecast Density}
For any point in state space, $\boldsymbol{x}$, and positive real number $\epsilon$, let $B_{\boldsymbol{x}}(\epsilon)$ denote an $\epsilon$-ball centred at $\boldsymbol{x}$. Suppose that $\varrho$ is some invariant measure (see appendix~\ref{app:inv}) associated with the attractor $A$. For any $\boldsymbol{x}_0\in A$, we define a new probability measure associated with $B_{\boldsymbol{x}_0}(\epsilon)$ by $\varrho_0(E)=\varrho(E\cap B_{\boldsymbol{x}_0}(\epsilon))/\varrho(B_{\boldsymbol{x}_0}(\epsilon))$. This measure induces some probability density function, $p_0(\boldsymbol{x};\boldsymbol{x}_0,\epsilon)$. We will call a set of points drawn from $p_0(\boldsymbol{x};\boldsymbol{x}_0,\epsilon)$ a {\it perfect initial ensemble}. At any time $t$, the forecast of the perfect initial ensemble using the flow $\boldsymbol{\varphi}_t$ will be distributed according to some pdf $p_t(\boldsymbol{x};\boldsymbol{x}_0,\epsilon)$. The pdf $p_t(\boldsymbol{x};\boldsymbol{x}_0,\epsilon)$ will be referred to as a {\it perfect forecast density} at lead time $t$.
\subsection{Imperfect Forecasts}
Operationally, we never get a perfect forecast since our initial ensemble is never drawn from $p_0(\boldsymbol{x},\boldsymbol{x}_0,\epsilon)$ and our model, $\boldsymbol{\varphi}_t(\boldsymbol{x})$, is always some approximation of the system, $\bar{\boldsymbol{\varphi}}_t(\boldsymbol{x})$, which possibly lives in a different state space. In that case, our forecast pdf would be $f_t(\boldsymbol{x};\boldsymbol{x}_0,\epsilon)$ rather than  the perfect forecast $p_t(\boldsymbol{x};\boldsymbol{x}_0,\epsilon)$. Indeed many, if not all, density forecasts issued in different applications arise from imperfect models. In particular, density forecasts from weather centres across the world and fan chart forecasts from the Bank of England arise from imperfect models. It was the observation that perfect density forecasts may be unattainable in practise that led Gneiting et al.~\cite{til-gne} to introduce a new paradigm of probabilistic forecasting. The lead author discusses this paradigm in another paper under review elsewhere. Suffice it to say that density forecasts from imperfect models represent our best estimate of the associated perfect forecasts in as much as imperfect models are our best representations of reality.
\section{Scoring Probabilistic Forecasts}
\label{sec:scor}
The next question would be: how close is $f_t(\boldsymbol{x};\boldsymbol{x}_0)$ to $p_t(\boldsymbol{x};\boldsymbol{x}_0)$? If one has a set of competing models, all of which may be imperfect, it might be of interest to rank the models and determine the best. These questions may be addressed using a scoring rule. In a general sense, we consider the score of a forecast $f_t(\boldsymbol{x};\boldsymbol{x}_{\tau})$ and denote it by $S(f_t(\boldsymbol{x};\boldsymbol{x}_{\tau}),\boldsymbol{X})$~\cite{joc-06}, where $\boldsymbol{X}$ is the random variable of which $\boldsymbol{x}$ is a particular realisation. If $\boldsymbol{X}$ is distributed according to $p_t(\boldsymbol{x};\boldsymbol{x}_{\tau})$, the expected score of $f_t$ is
\begin{equation}
\mathbb{E}[S(f_t(\boldsymbol{x};\boldsymbol{x}_{\tau}),\boldsymbol{X})]=\int S(f_t(\boldsymbol{x};\boldsymbol{x}_{\tau}),\boldsymbol{z})p_t(\boldsymbol{z};\boldsymbol{x}_{\tau})\ud\boldsymbol{z}.
\label{s1:eq7}
\end{equation}

At lead time, $t$, the overall forecast score on the attractor is
\begin{equation}
\mathbb{E}[S(t)]=\lim_{T\rightarrow\infty}\frac{1}{T}\int_0^T\mathbb{E}[S(f_t(\boldsymbol{x};\boldsymbol{x}_{\tau}),\boldsymbol{X}^{(\tau,t)})]\ud\tau,
\label{s1:eq8}
\end{equation}
where $\boldsymbol{X}^{(\tau,t)}$ is the random variable being forecast from the initial distribution corresponding to $\boldsymbol{x}_{\tau}$. Provided the underlying attractor is ergodic, we can rewrite~(\ref{s1:eq8}) as
\begin{equation}
\mathbb{E}[S(t)]=\lim_{T\rightarrow\infty}\frac{1}{T}\int_0^TS(f_t(\boldsymbol{x};\boldsymbol{x}_{\tau}),\boldsymbol{X}^{(\tau,t)})\ud\tau.
\label{s1:eq9}
\end{equation}
In all practical situations, corresponding to each density forecast will be one observation. The distribution $p_t(\boldsymbol{x};\boldsymbol{x}_0)$ is never available for use in the evaluation. It follows that we cannot know how close a single forecast distribution is to the target distribution. The best we can do is to evaluate a forecasting system over a collection of forecasts made from different initial states. Since the target distribution is not available in practise, we can use~(\ref{s1:eq9}) to score forecast performance rather than~(\ref{s1:eq8}). When used this way, one can rank competing models or assess the goodness of a given model. Scores of probabilistic forecasts are also useful for estimating parameters~\cite{til-07}.

Let us now discretise time according to $\tau_i=(i-1)\tau_s$, for $i=1,2,..,N$, where $\tau_s$ is the sampling time. This gives a sequence of forecast pdfs, $\{f_t(\boldsymbol{x};\boldsymbol{x}_i)\}_{i=1}^N$, corresponding to verifications $\{\boldsymbol{X}^{(i,t)}\}_{i=1}^N$ and score $S$. We can thus discretise~(\ref{s1:eq9}) to obtain the following empirical score to value the forecast system at lead time $t$:
\begin{equation}
\langle S\rangle(t)=\frac{1}{N}\sum_{i=1}^NS(f_t(\boldsymbol{x};\boldsymbol{x}_i),\boldsymbol{X}^{(i,t)}).
\label{s1:eq10}
\end{equation}
This is the same score proposed by Brocker and Smith~\cite{joc-06}. The associated discretisation error in moving from~(\ref{s1:eq9}) to (\ref{s1:eq10}) is 
\begin{equation*}
\mbox{\huge{$\varepsilon$}}\le\frac{T*\tau_s^2}{24}\times\frac{d^2S}{d\tau^2}(f_t(\boldsymbol{x};\boldsymbol{x}_{\vartheta}),\boldsymbol{X}^{(\vartheta,t)}),
\end{equation*}
where $\vartheta\in(0,T)$ and $S$ is twice continuously differentiable. This error vanishes as $N\rightarrow\infty$. Further to that, it is important for the observation time $T$ to be large to ensure that the observations sample underlying invariant distribution well.

In this paper, we shall use the Ignorance score:
\begin{equation*}
S(f_t,\boldsymbol{X})=\mbox{ign}(f_t,\boldsymbol{X}),
\end{equation*}
where $\mbox{ign}(f_t,\boldsymbol{X})=-\log f_t(\boldsymbol{X})$ is the logarithmic scoring rule originally proposed by Good~\cite{good-52}. It measures the amount of information contained in the forecast about the system in question. Therefore, it is ``the information deficit, or {\it Ignorance}'' that a forecaster in possession of the pdf has before making the observation $\boldsymbol{X}$~\cite{roul-02}. An important property of this score is that it is {\it strictly proper}. A strictly proper score is one for which~(\ref{s1:eq7}) assumes its minimum if and only if $f_t=p_t$~\cite{til-07}. Another property of the Ignorance score, although less persuasive, is {\it locality}. A score is local if it only requires the value of the forecast pdf at the verification to be evaluated~\cite{joc-06}. Bernardo~\cite{ber-79} proved that the Ignorance score is the only proper score that is also local. Nonetheless, being proper seems more important than locality. In fact it has been demonstrated in~\cite{til-07} that to use a score that is not proper is ill-advised. Another attractive property of the Ignorance score is that under some betting scenario, it represents the rate at which a forecaster's wealth grows with time~\cite{joc-07,hag-08}. A common objection to using the Ignorance score is that it gives a heavy penalty to issuing low probabilities to events that do happen. For instance, Gneiting and Raftery~\cite{til-07} found it to yield higher estimates of forecast spreads than other proper scoring rules. Brocker and Smith~\cite{joc-07} argue that the Ignorance score yields larger parameter estimates as a sign that there are bad forecasts which need to be dealt with appropriately. Dealing with such bad forecasts is especially important when the score to use is not a matter of choice. Moreover, bad forecasts are inevitable whenever there is model error. The next section discusses a way to deal with bad forecasts. 
\section{Density-Forecast Estimation}
\label{sec:dfe}
In this section, we consider how to move from discrete to continuous forecasts. Suppose we have an ensemble of discrete forecasts, $\{Y_j^{(\tau,t)}\}_{j=1}^M$, at lead time $t$ and with corresponding initial condition at time $\tau$. We convert these into a density forecast by using kernels, $K(\eta)$. A possible density estimate of this ensemble is
\begin{equation}
\rho_{t}^{(\tau)}(x)=\frac{1}{\sigma_k^{(t)}M}\sum_{j=1}^MK\left\{\left(x-Y_j^{(\tau,t)}-\mu^{(t)}\right)/\sigma_k^{(t)}\right\},
\label{dfe:eq1}
\end{equation}
where $\sigma_k^{(t)}$ is the kernel width and $\mu^{(t)}$ is the offset parameter, which takes into account possible bias in the ensembles that may arise due to model error. This parameter makes~(\ref{dfe:eq1}) differ from Parzen's~\cite{par-62} density estimates. Note that we cannot follow Silverman~\cite{sil-86} to select the kernel width because such an approach does not account for model error. To select the parameters, we appeal to the scoring rule discussed in the previous section, using an archive of pairs of ensembles and corresponding verifications. But there is still a problem. If the ensembles tend to be far from the verifications, then the kernel width chosen may be too wide thus reducing the value of the density forecasts~\footnote{The lead author discusses this further in another paper currently under review elsewhere.}. This problem is a result of model error.

To circumvent the model error problem, we can blend with climatology as suggested by~\cite{joc-07} to obtain:
\begin{equation}
f_t^{(\tau)}(x)=\alpha\rho_t^{(\tau)}(x)+(1-\alpha^{(t)})\rho(x),
\label{dfe:eq2}
\end{equation}
where $\rho(x)$ is the climatology (or invariant distribution), estimated from an historical data archive using kernels. The parameters $\alpha^{(t)}$, $\mu^{(t)}$ and $\sigma_k^{(t)}$ are selected simultaneously by minimising the Ignorance score. When there is model error, values of $\alpha^{(t)}\neq1$ and $\mu^{(t)}\neq0$ will be selected, ensuring that $\sigma_k^{(t)}$ is not chosen too large. In effect, the density forecasts issued will express more confidence. A value of $\alpha^{(t)}=0$ indicates that the model has no forecasting value at the given lead time. Hence the Ignorance score of the climatology is a bench mark for any given model.

In practise, we will consider forecasts given at discrete time steps, $\tau_i=(i-1)\tau_s$, $i=1,2,\ldots,N$ where $\tau_s$ is the sampling time. For a given lead, using Gaussian kernels, the parameters are then fitted by minimising
\begin{equation*}
\langle\mbox{ign}\rangle=-\frac{1}{N}\sum_{i=1}^N\log f_t^{(i)}(s_{i+t}),
\end{equation*}
where $\{s_i\}$ are given observations. Minimising the above score is equivalent to {\em quasi-maximum likelihood} under model error with independent conditional forecasts as discussed by White~\cite{whi-94}. Here, we do not assume that the forecasts are independent. In the graphs of \S~\ref{sec:init}, the values of the average Ignorance score are computed after the minimisation has been done.
\section{Radial Basis Function Models}
\label{sec:rbf}
All imperfect models considered in this paper are constructed from data using radial basis functions. A radial basis function is a function $\psi(\boldsymbol{x})$ such that $\psi(\boldsymbol{x})=\psi(|\boldsymbol{x}|)$, where $|\cdot|$ is some norm. We use the Euclidean norm throughout this paper. We consider the case where one is given only a scalar time series, $\{s_i\}_{i=1}^N$. One can then form delay vectors, $\boldsymbol{x}_i=(s_i,s_{i-\nu_d},\ldots,s_{i-(m-1)\nu_d})$, where $m$ is the embedding dimension and $\nu_d$ is some time delay. The time delay may be selected according to~\cite{fra-swi} and the embedding dimension by the method of false nearest neighbours~\cite{rho-97}.

For a given delay vector, $\boldsymbol{x}$, we construct the model $\phi(\boldsymbol{x}): \mathbb{R}^m\rightarrow\mathbb{R}$, which takes the form
\begin{equation*}
\phi(\boldsymbol{x})=\sum_{j=1}^{n_c}\lambda_j\psi(\boldsymbol{x}-\boldsymbol{c}_j)
\end{equation*}
where $\lambda_j$ are constants and $\boldsymbol{c}_j$ are centres, each determined to satisfy the condition $\phi(\boldsymbol{x}_i)=s_{i+1}$ as discussed in~\cite{jud-95}. One would then have the deterministic model
\begin{equation}
\hat{s}_{i+1}=\phi(\boldsymbol{x}_{i}),
\label{rbf:eq1}
\end{equation}
where $\hat{s}_{i+1}$ is a forecast of $s_{i+1}$.

The above model will not be a perfect representation of the underlying dynamics, not less because the functional form is not the correct form of the system dynamics. Depending on the system under consideration and the amount of data available, its forecasting performance can vary. When selecting the initial distribution spread, the inadequacies of this model would be accounted for by the methodology discussed in the previous section.

At this point we note that it has been argued in~\cite{jud-07} that model error cannot be adequately accounted for without including dynamical noise. While this may seem to undermine the value of our contribution, one should note a number of points concerning the example used in~\cite{jud-07}: (1) Raw ensembles were being evaluated; (2) the score used was a bounding box and (3) there was no observational error. We shall revisit this issue in \S~\ref{subsec:imp}, considering dynamical noise $\omega_i$ such that the model becomes 
\begin{equation}
\hat{s}_{i+1}=\phi(\boldsymbol{x}_{i})+\omega_i.
\label{rbf:eq2}
\end{equation}
We will discuss the effect of observational error on the value of this model approach and, as a byproduct of the discussion, provide a way for estimating the spread of $\omega_i$. It will be assumed that $\mathbb{E}[\omega_i]=0$ and $\mathbb{E}[\omega_i^2]=\sigma_{\omega}^2$.
\section{Initial Distribution Spread}
\label{sec:init}
The primary concern is to determine optimal initial distribution spreads for the forecasting problem. Each initial ensemble is drawn from a Gaussian distribution centred at the initial observation. The problem is then reduced to finding the optimal spread of the Gaussian distribution. An optimal spread is one that minimises the average score at the lead time of interest. In the theoretical setup this score would be the one given by equation~(\ref{s1:eq8}) and in an operational setup we would use the empirical score given by~(\ref{s1:eq10}). In the numerical examples considered in this section, we use continuous forecast pdfs obtained from discrete forecasts as discussed in \S~\ref{sec:dfe}. The average Ignorance score of the forecast pdfs is given relative to that of the climatology.

The cases considered are the perfect model scenario and the imperfect model scenario. In the perfect model scenario, numerical experiments are performed on the Moore-Spiegel system~\cite{mor-spi} at classical parameter values. In the imperfect model scenario, the Moore-Spiegel system and circuit are considered and the models are constructed from data using cubic radial basis functions as discussed in \S~\ref{sec:rbf}~(see~\cite{jud-95,res-dphil} for more details). We shall denote the spread of the underlying Gaussian distribution of the initial ensemble by $\sigma_e$, that of the dynamical noise by $\sigma_{\omega}$ and that of the observational error by $\delta$. For observational error of a given spread, we vary $\sigma_e$ (or $\sigma_{\omega}$) logarithmically between $10^{-3}$ and 1. $\delta=0$ will represent the noise-free case. In the multivariate case, we set $\sigma_e$ to be the spread of the perturbation of the $i$th coordinate and then set the spread of the perturbation of the $j$th coordinate to be $\sigma_e^{(j)} = \sigma_e*\sigma_j/\sigma_i$, where $\sigma_j$ is the standard deviation of the $j$th variable. Which coordinate is chosen to be the $i$th coordinate does not matter. In the subsequent discussions, we will have three coordinates $(x,y,z)^T$ and we select $z$ to be the $i$th variable with $\delta$ as the spread of observational error on this variable. The spread of observational error on the $j$ coordinate is then set to be $\delta_j=\delta*\sigma_j/\sigma_i$. This constrains the signal to noise ratio to be the same on all the variables. Note that we assume the noise to be spatially uncorrelated.
\subsection{Perfect Model Scenario}
We consider the Moore-Spiegel system~\cite{mor-spi}, which is given by:
\begin{equation}
\begin{array}{lll}
\dot{x}&=&y,\\
\dot{y}&=&-y+R x-\Gamma(x+z)-R xz^2,\\
\dot{z}&=&x,
\end{array}
\label{s2:eq1}
\end{equation}
with classical parameters $\Gamma\in[0,50]$ and $R=100$. This system was integrated for $\Gamma=36$ and $R=100$ using a 4-th order Runge-Kutta method to generate some data, which we will call {\it Moore-Spiegel data}. The maximal Lyapunov exponent for this system was found to be $\sim 0.22$, which corresponds to a doubling time of $\approx1.43$. The integration time step used was 0.01, but we sampled every 4 points. Thus the time step between consecutive data points collected was $\delta t=0.04$. Transients were discarded to ensure that all the data collected were very close to the attractor. 
From any initial point on the Moore-Spiegel data, an initial ensemble is generated by perturbing the observation with some random variable drawn from a Gaussian distribution with the Moore-Spiegel system in~(\ref{s2:eq1}) used as the model to make forecast distribution.
\subsubsection{Clean Data}
\begin{figure}[!t]
\centering
\includegraphics[height=7.5cm,width=10cm]{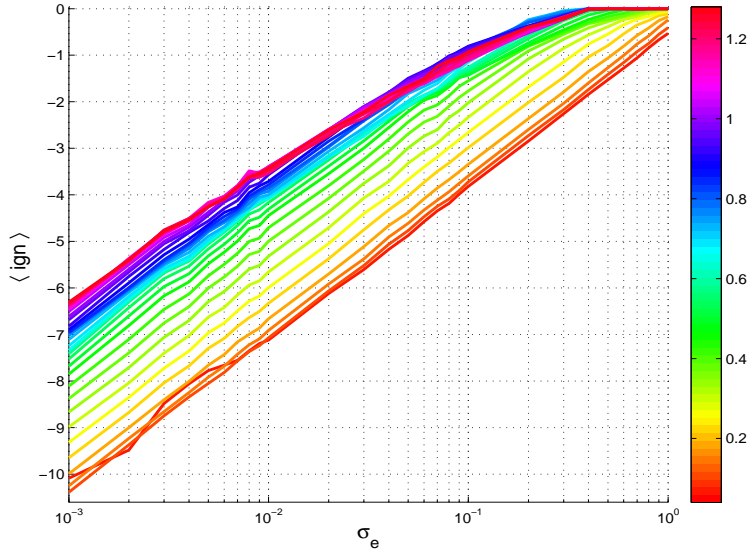}
\caption{\small Graphs of the average Ignorance score versus initial distribution spread, $\sigma_e$, using a perfect Moore-Spiegel model with 512 ensembles for various lead times (according to the right colour bar), each ensemble containing 32 members. The Moore-Spiegel data was noise-free.} 
\label{pf01}
\end{figure}

For the case $\delta=0$, graphs of the average Ignorance score, $\langle\mbox{ign}\rangle$ , versus the initial distribution spread, $\sigma_e$, are shown in figure~\ref{pf01}.
 The different colours correspond to the different lead times of up to 32 time steps, each time step being $\delta t=0.04$. Notice that the graphs generally yield straight lines except at higher lead times and initial distribution spread. In particular, the magenta lines (corresponding to lead times of 32$\delta t$) saturate at higher values of $\sigma_e$. As the initial distribution spread increases, we would expect the forecast pdfs at low lead times to be approximately flattened Gaussians. That is why all the red lines (lead times 1 and 2) grow linearly without saturating. Notice that the lead times of, say 31$\delta t$ and 32$\delta t$, score less than some lower lead times when $\sigma_e>10^{-1}$. When higher lead time forecasts score less than lower lead times, we say we have {\it return of skill}~\cite{bar-07}. Linear graphs such as those in figure~\ref{pf01} suggest that the underlying model is perfect and the data is noise-free.
\subsubsection{Noisy Data}
\begin{figure}[!t]
\hspace{-0.6cm}
\includegraphics[width=7.5cm,height=7.5cm]{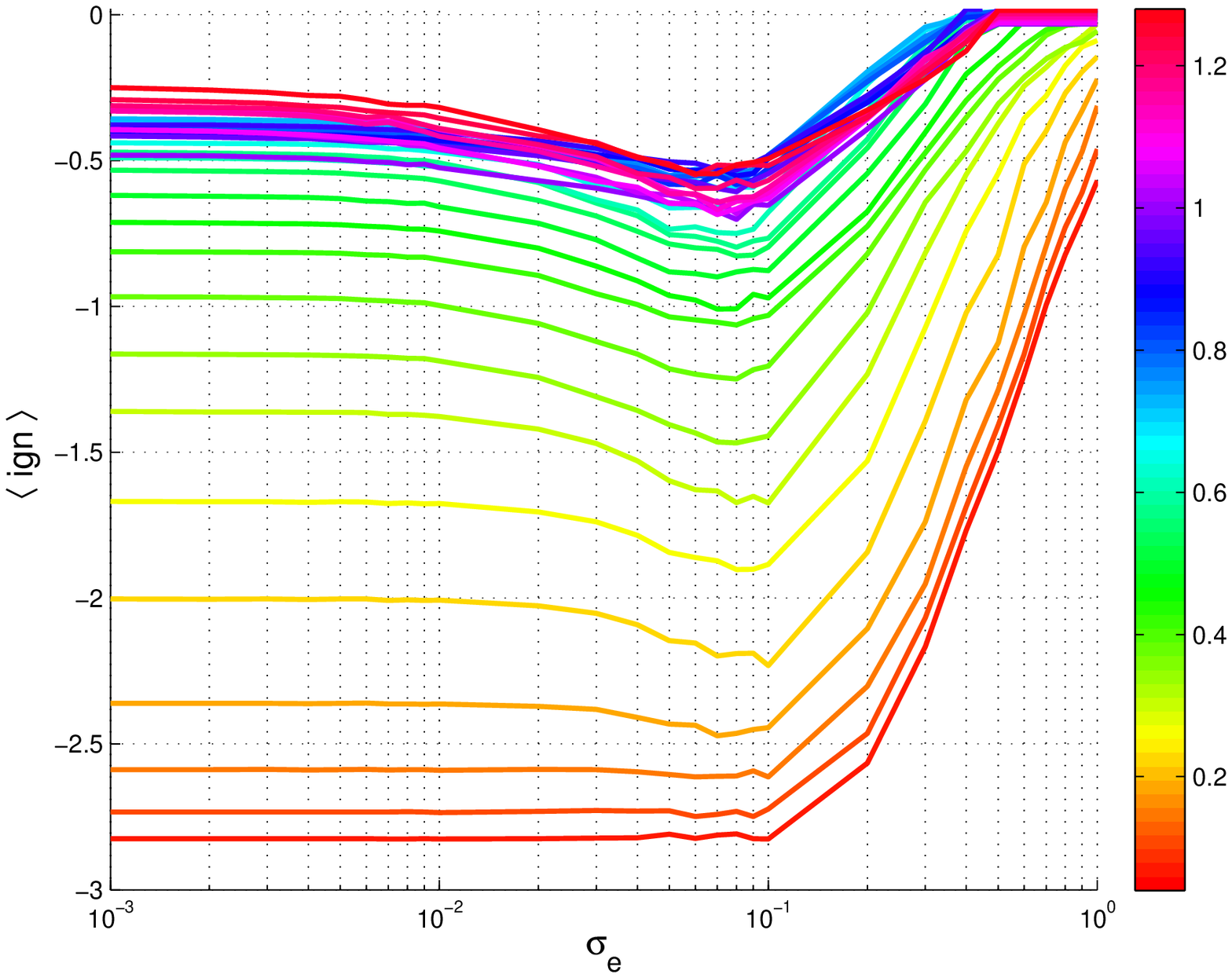}
\includegraphics[width=7.5cm,height=7.5cm]{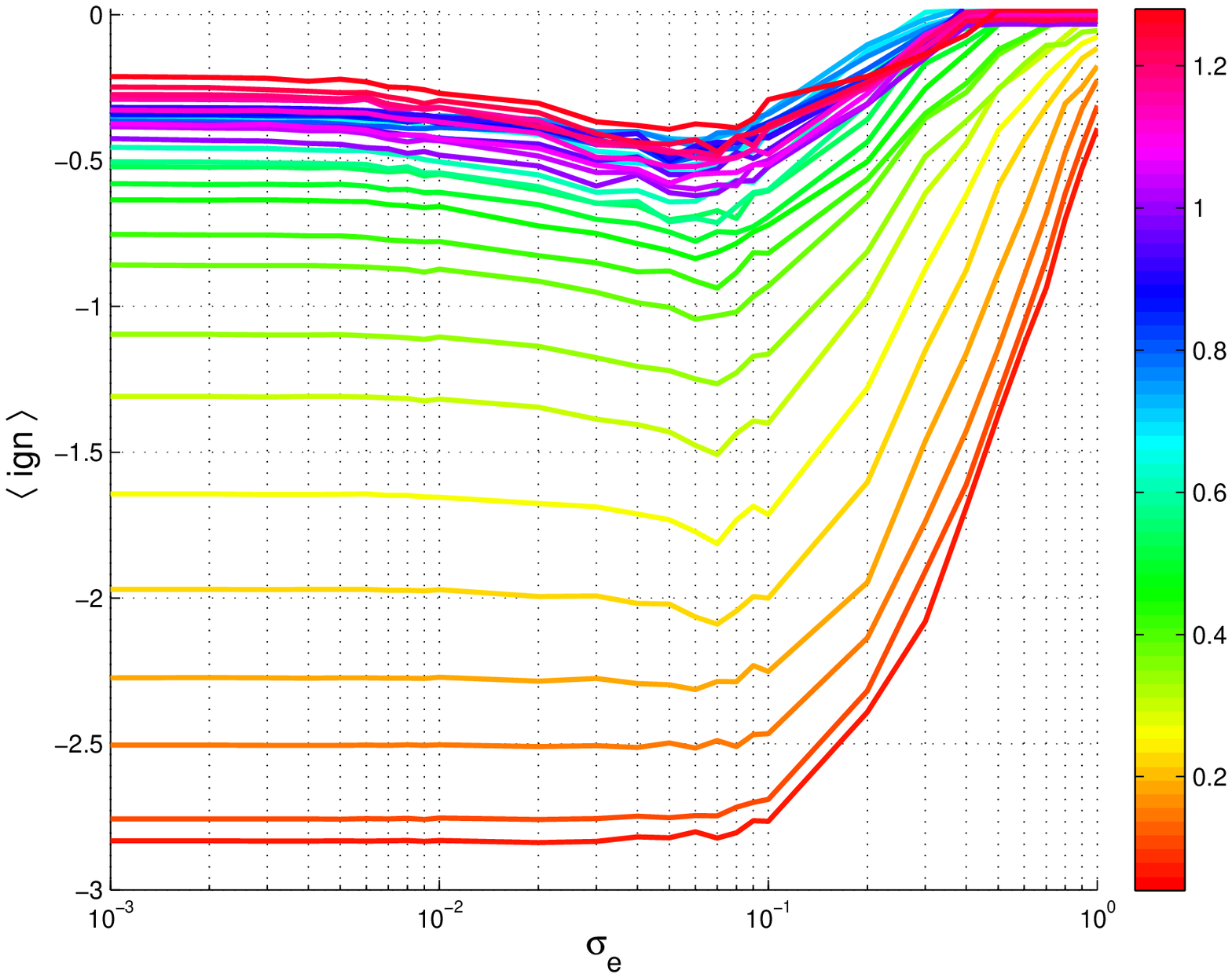}
\caption{\small The graph of average Ignorance score versus initial distribution spread, $\sigma_e$, for the perfect Moore-Spiegel model with 512 ensembles for various lead times (according to the right colour bar), each ensemble containing 32 members (left) and 9 members (right). Observational error was $\delta=10^{-1}$.}
\label{pf1}
\end{figure}

\begin{figure}[!t]
\centering
\hbox{
\includegraphics[width=0.5\textwidth]{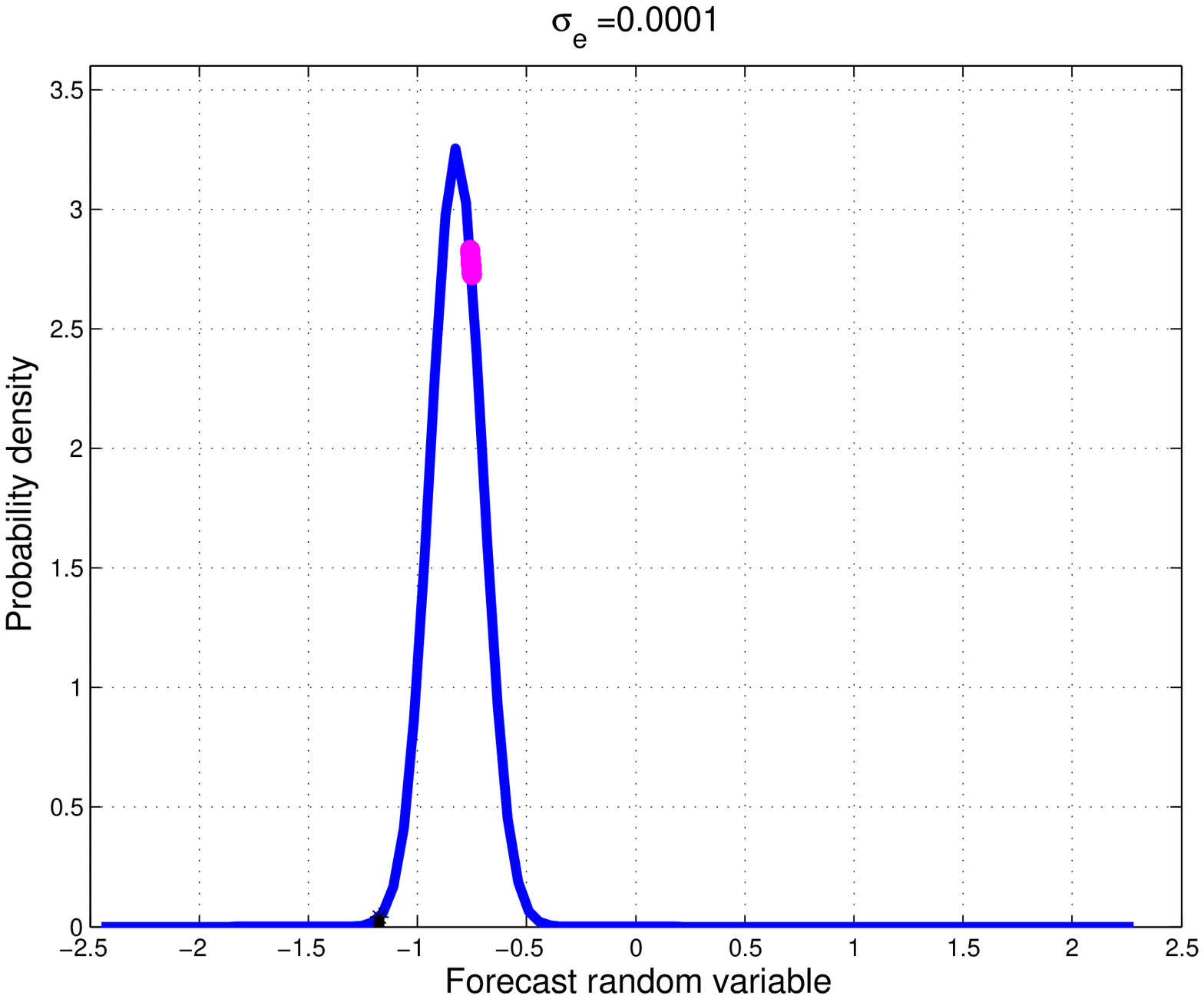}
\includegraphics[width=0.5\textwidth]{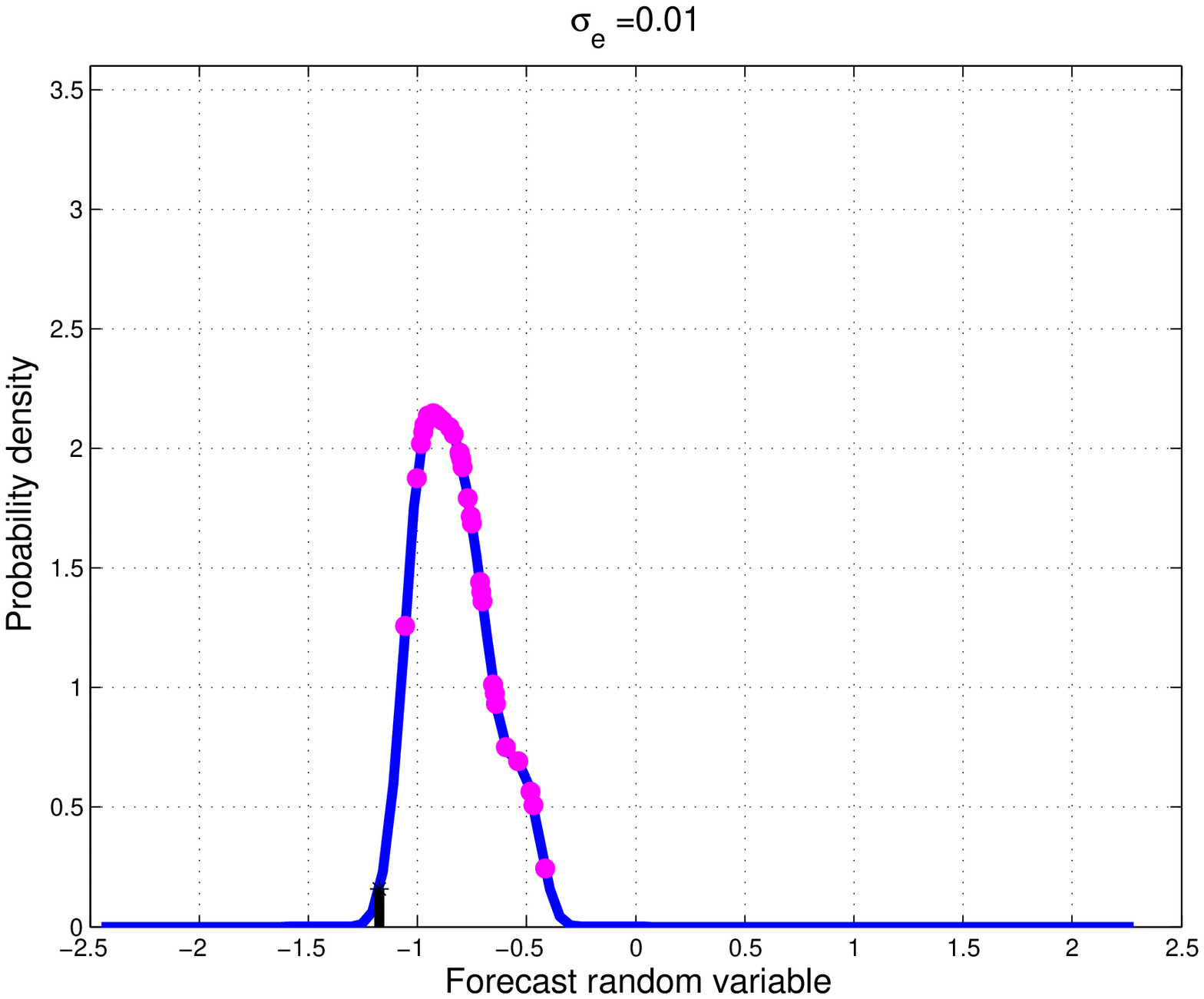}}
\hbox{
\includegraphics[width=0.5\textwidth]{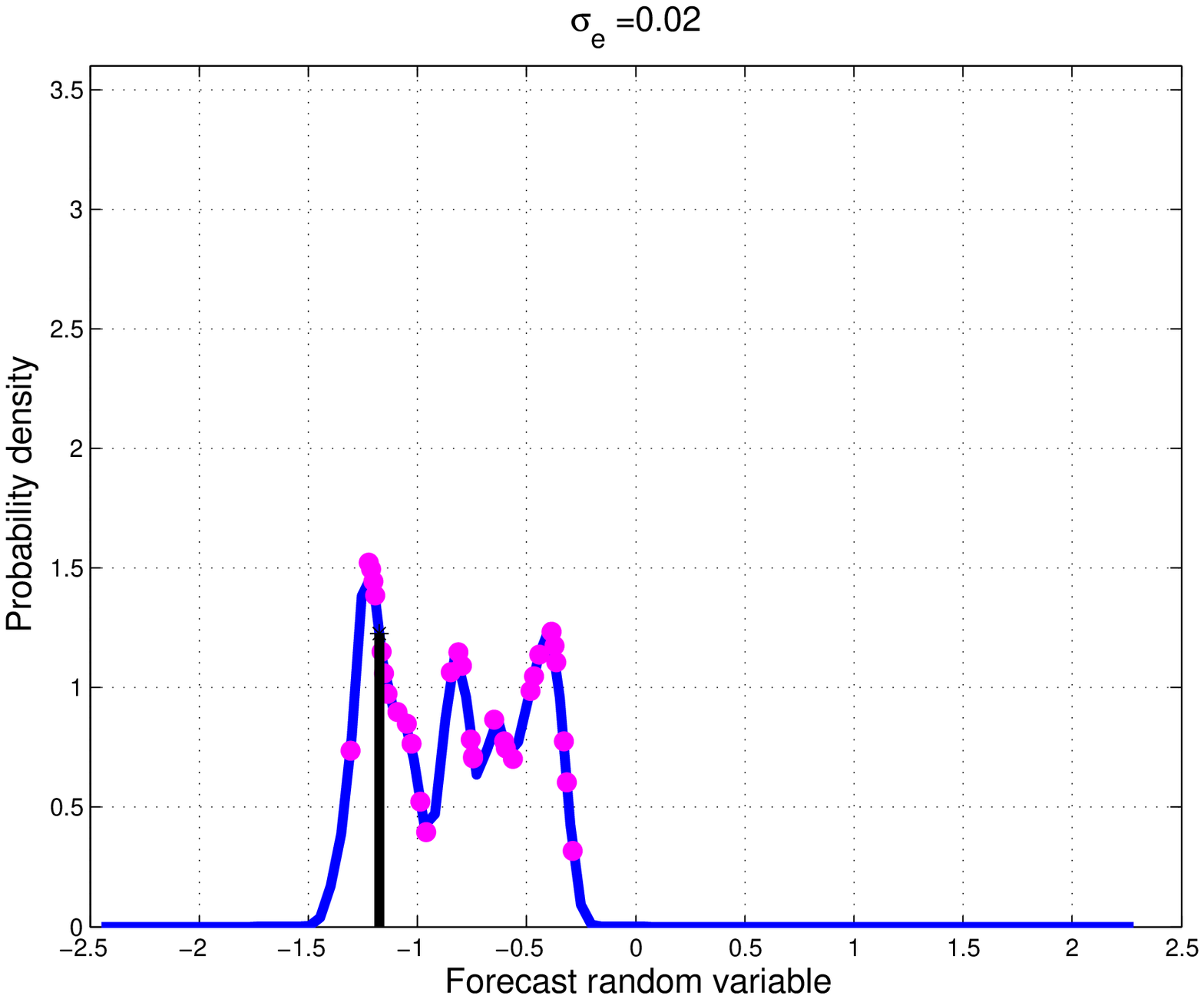}
\includegraphics[width=0.5\textwidth]{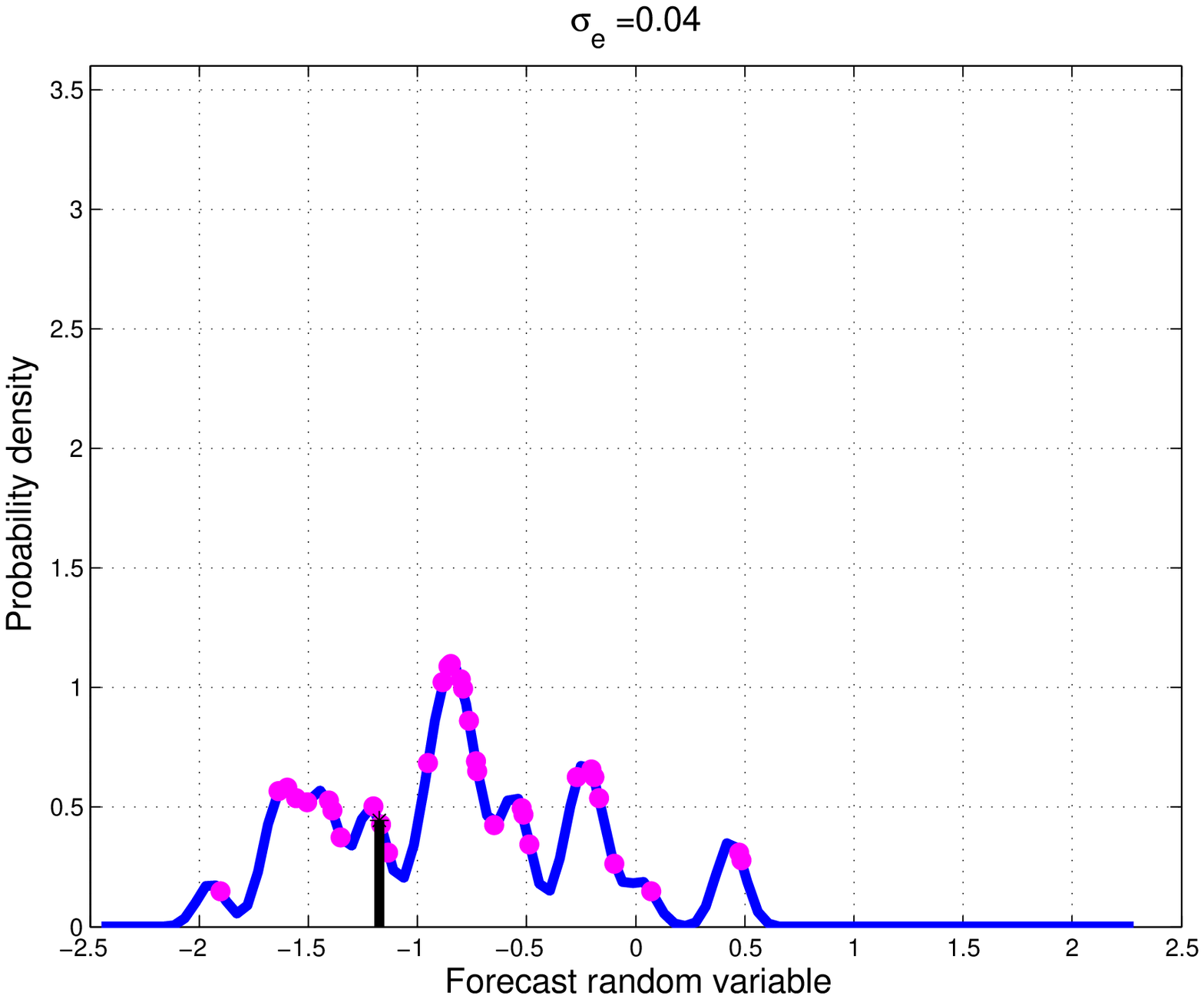}}
\caption{\small A series of ensemble pdfs at a lead time of 16 time-steps from the initial condition. The magenta dots are the actual ensemble members and the blue line is the fitted distribution. Notice that as the initial distribution spread, $\sigma_e$, increases, the value of $f_t(x^*)$ (indicated by the black vertical line) increases and then decreases, where $x^*$ is the verification.}
\label{pdf_series}
\end{figure}

We next consider the case when the data has observational error of standard deviation, $\delta=10^{-1}$ (i.e. $\delta_j/\sigma_j\approx 0.1$). The corresponding graphs of the average Ignorance score versus  the initial distribution spread, $\sigma_e$, are shown in figure~\ref{pf1}.
At low initial distribution spread, all the graphs are almost flat since the initial distribution spreads are drowned by the noise. As the initial distribution spread increases, the higher lead time graphs begin to dip. This is because the forecast pdfs at higher lead times spread out, and in the process, the verifications which were initially at the tails of the distributions, tend to be encapsulated by the ensembles as we gain skill (see figure~\ref{pdf_series}).
At low lead times, the verifications are generally at the centre of the ensembles. As the distributions spread out and flatten, the average Ignorance score increases. That is why at low lead times, the graphs, initially flat, begin to increase linearly. This occurs when $\sigma_e\approx\delta$: the same value of $\sigma_e$ at which graphs of the higher lead times attain their minima. Graphs of the average Ignorance score become horizontal when $\sigma_e\ge 4*10^{-1}$. This is where performance matches that of climatology.

In figure~\ref{pf1}, the graphs on the left were obtained with 32 members per ensemble and those on the right with 9 members. A 9 member ensemble would provide a poor estimate of an underlying distribution. The similarity between both graphs is an indication that the method is robust to sampling errors. That the ensemble size 9 is poorer is reflected by higher average Ignorance values on the right hand side panel of figure~\ref{pf1}. Nonetheless, in both cases predictability is not lost within all the lead times considered for all $\sigma_e\le 10^{-1}$.
\subsection{Imperfect Model Scenario}
\label{subsec:imp}
Let us now turn to the imperfect model scenario. The models were constructed using radial basis functions as discussed in \S~\ref{sec:rbf}. We start by comparing models~(\ref{rbf:eq1}) and~(\ref{rbf:eq2}) to assess the value of dynamical noise relative to perturbing the initial conditions to account for model error. We first consider the Moore-Spiegel system and then move on to an electronic circuit.
\subsubsection{The Moore-Spiegel system}
Data was generated by integrating the Moore-Spiegel system with an integration time step of 0.01. Transients were first eliminated to ensure the dynamics were near the attractor. The data was then sampled every fourth time step so that the time interval between successive points in the new data set was $\delta t=0.04$. The new data set is what we work with henceforth. A radial basis function model of the form~(\ref{rbf:eq1}) was fitted on $10^4$ data points, with delay vectors formed on the $z$ variable. The time delay was chosen to be four time steps, the delay space had dimension $m=3$ and the number of centres used was $n_c=40$. 

\begin{figure}
\hspace{-0.6cm}
\includegraphics[width=7.5cm,height=7.5cm]{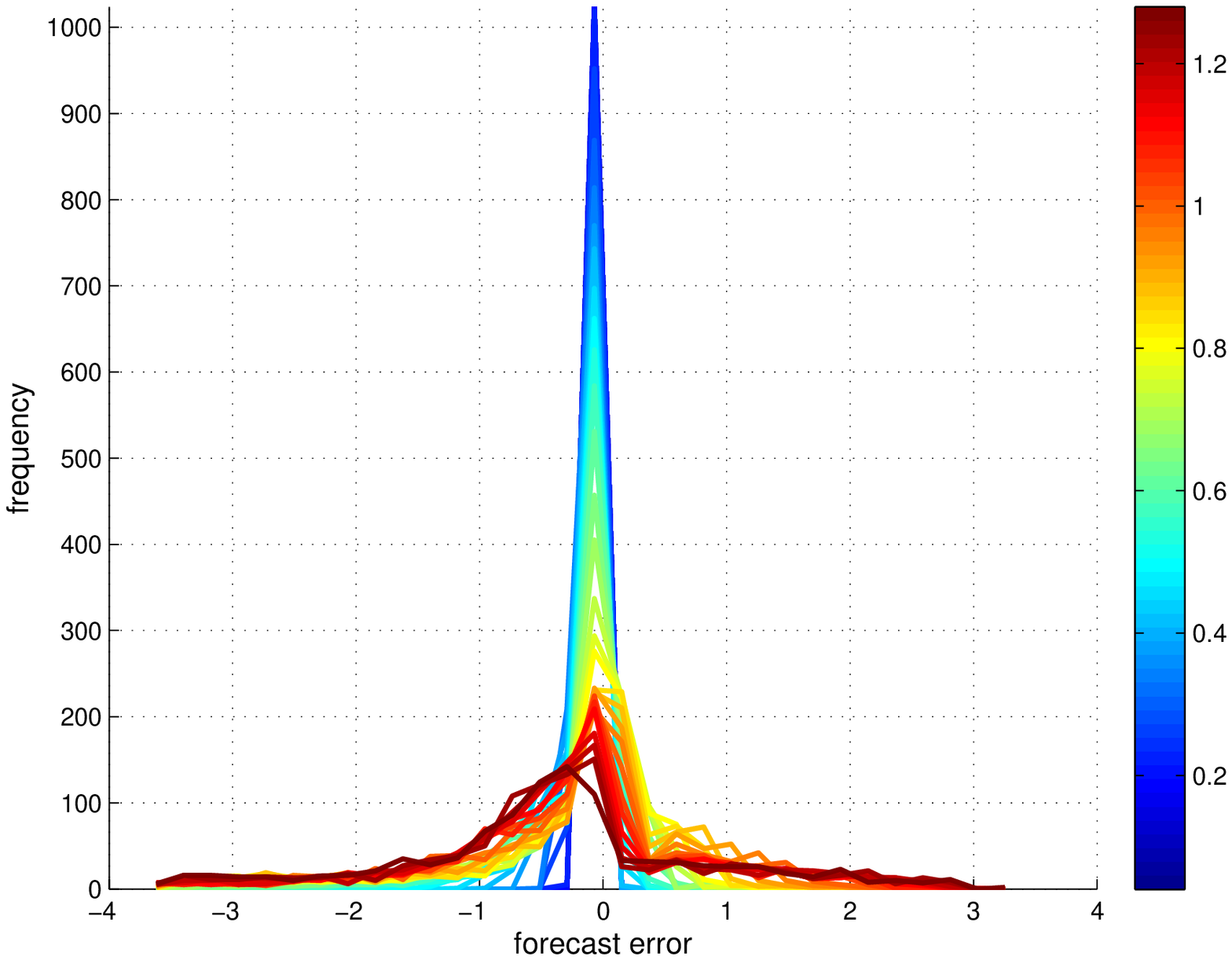}
\includegraphics[width=7.5cm,height=7.5cm]{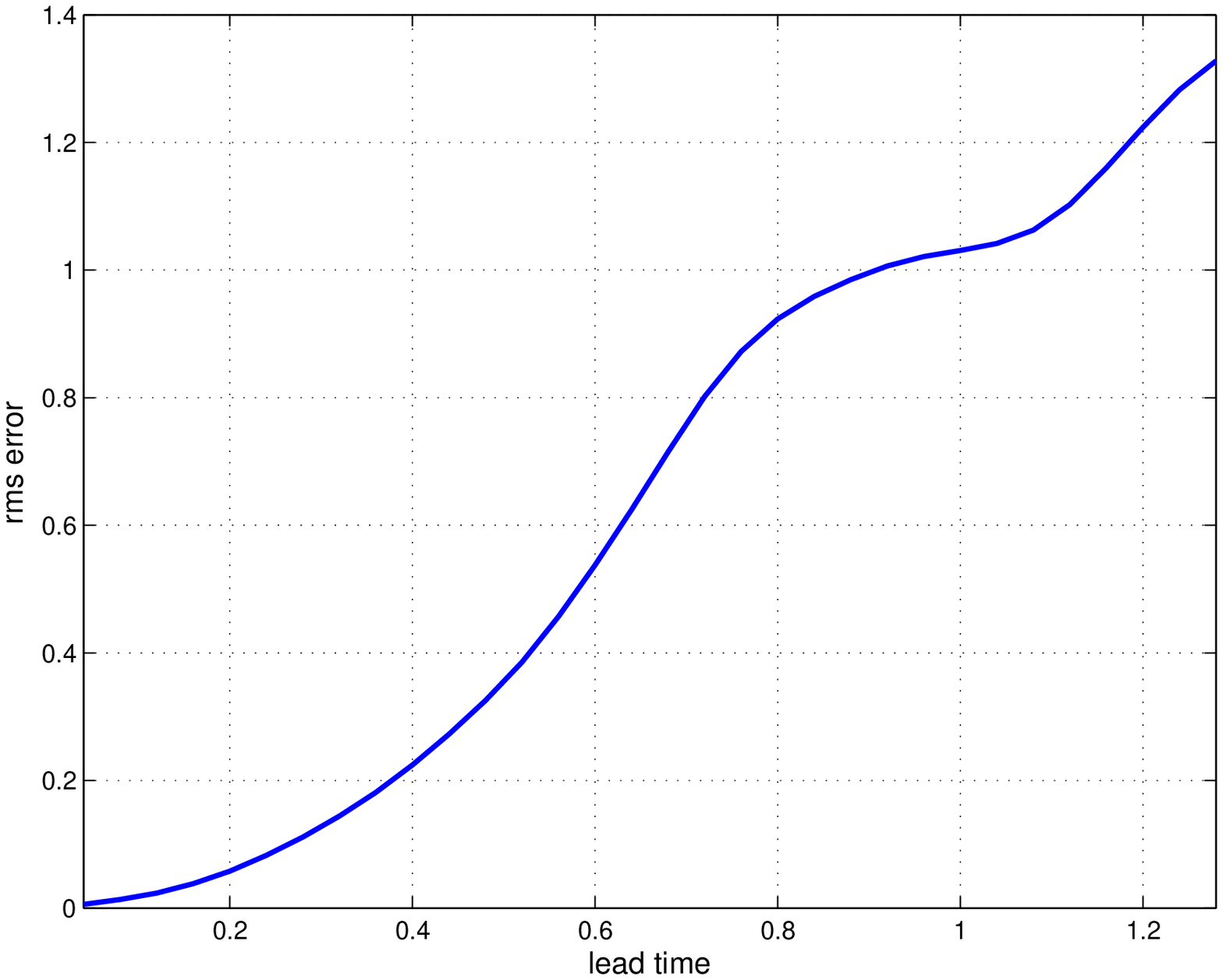}
\caption{(left) Distributions of forecast errors for the $z$ variable and various lead times as indicated by the colorbar  and (right) a graph of root mean square of forecast errors as a function of lead time.}
\label{imp:fig1}
\end{figure}

Our first aim is to quantify model error in the absence of observational error and discuss how to mitigate it. Distributions of forecast errors up to lead time of $32\times\delta t$ for this model are shown in figure~\ref{imp:fig1}. As one would expect, the errors to grow with lead time. If we bear in mind that the data is noise-free, these forecast errors are evidence of model error. The graph of root mean square error of forecasts for lead times up to 32 time steps is shown  on the right hand side of the same figure. This gives additional insights into the extent of model error as a function of lead time. The standard deviation of the $z$ variable is about 1.12. According to the graph, this value is exceeded after a lead time of about 1.1 seconds (or $27\times\delta t$). We will soon see that density forecasts can still be useful beyond this lead time.

\begin{figure}[!t]
\hspace{-0.6cm}
\hbox{
\includegraphics[width=7.5cm,height=7.5cm]{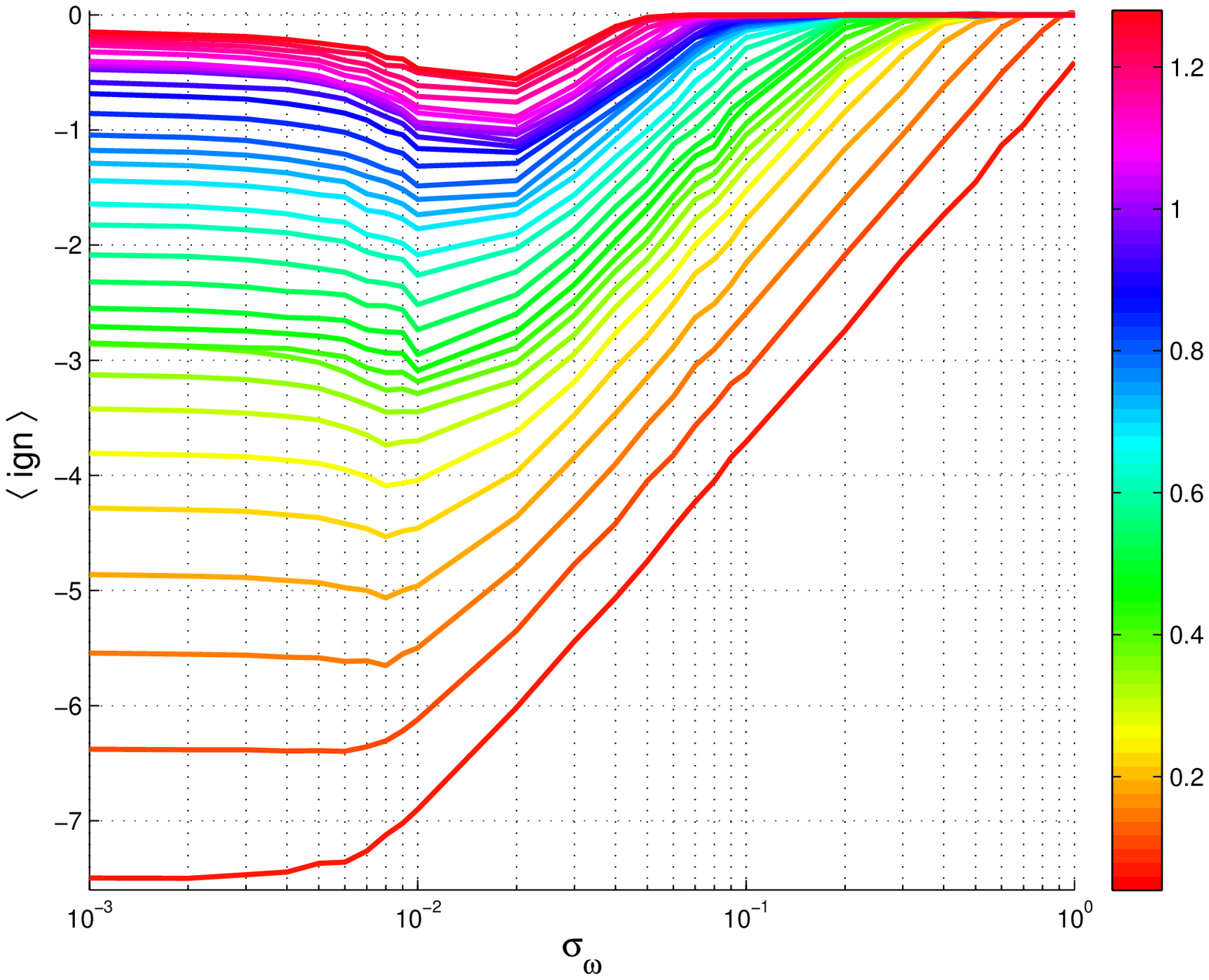},
\includegraphics[width=7.5cm,height=7.5cm]{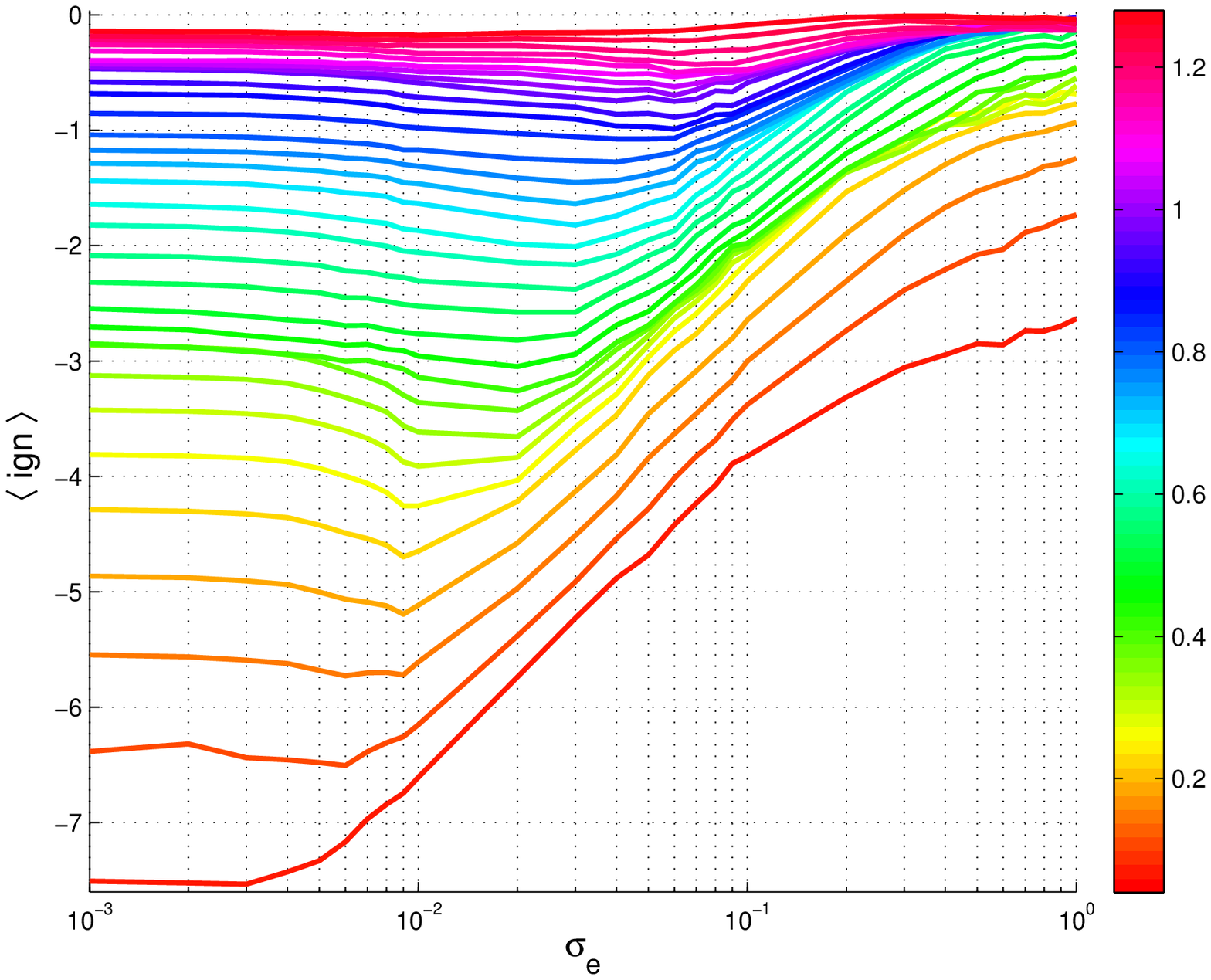}
}
\caption{\small The graphs of average Ignorance score versus dynamical noise (left) and initial distribution spread (right) with 512 ensembles, each ensemble containing 32 members and using a cubic radial basis function model on noise-free Moore-Spiegel data. The colour bar on the right shows the lead times for the different graphs of the Ignorance score.}
\label{jan3:e0}
\end{figure}

With the data still noise-free, we explore the possibility of mitigating model error by either the initial distribution spread with model~(\ref{rbf:eq1}) or by dynamical noise according to equation~(\ref{rbf:eq2}). At various lead times, performance of density forecasts as a function of the spread of dynamical noise is depicted in figure~\ref{jan3:e0} on the left. In this case the dynamical noise model is run without perturbations in the initial conditions. The dynamical noise is drawn from a Gaussian distribution with mean 0 and variance $\sigma_{\omega}^2$. Notice that graphs of the average Ignorance score dip at $\sigma_{\omega}=\sigma_{\omega}^*\approx10^{-2}$ for higher lead times and rise at approximately the same time for lower lead times. These graphs are very similar to those obtained in the perfect model scenario with observational error. If one opted to use the model with dynamical noise, then the best perturbations' spread to use is $\sigma_{\omega}^*=10^{-2}$. Empirical results of performance of density forecasts as a function of the initial distribution spread are shown in figure~\ref{jan3:e0} on the right. They are qualitatively similar to the perfect model scenario with noisy data. 
This case also presents striking differences from the previous noisy data scenario and the noise-free, dynamical noise case just discussed in that the optimal spread varies with lead time. In this case, one may select the initial distribution spread that yields good forecasting performance at higher lead times. Evidently, the dynamical noise model is superior at higher lead times where performance is close to that of climatology. On the other hand, perturbing the initial conditions and using the model with no dynamical noise is superior at lower lead times. 

\begin{figure}[!t]
\hspace{-0.6cm}
\includegraphics[width=7.5cm,height=7.5cm]{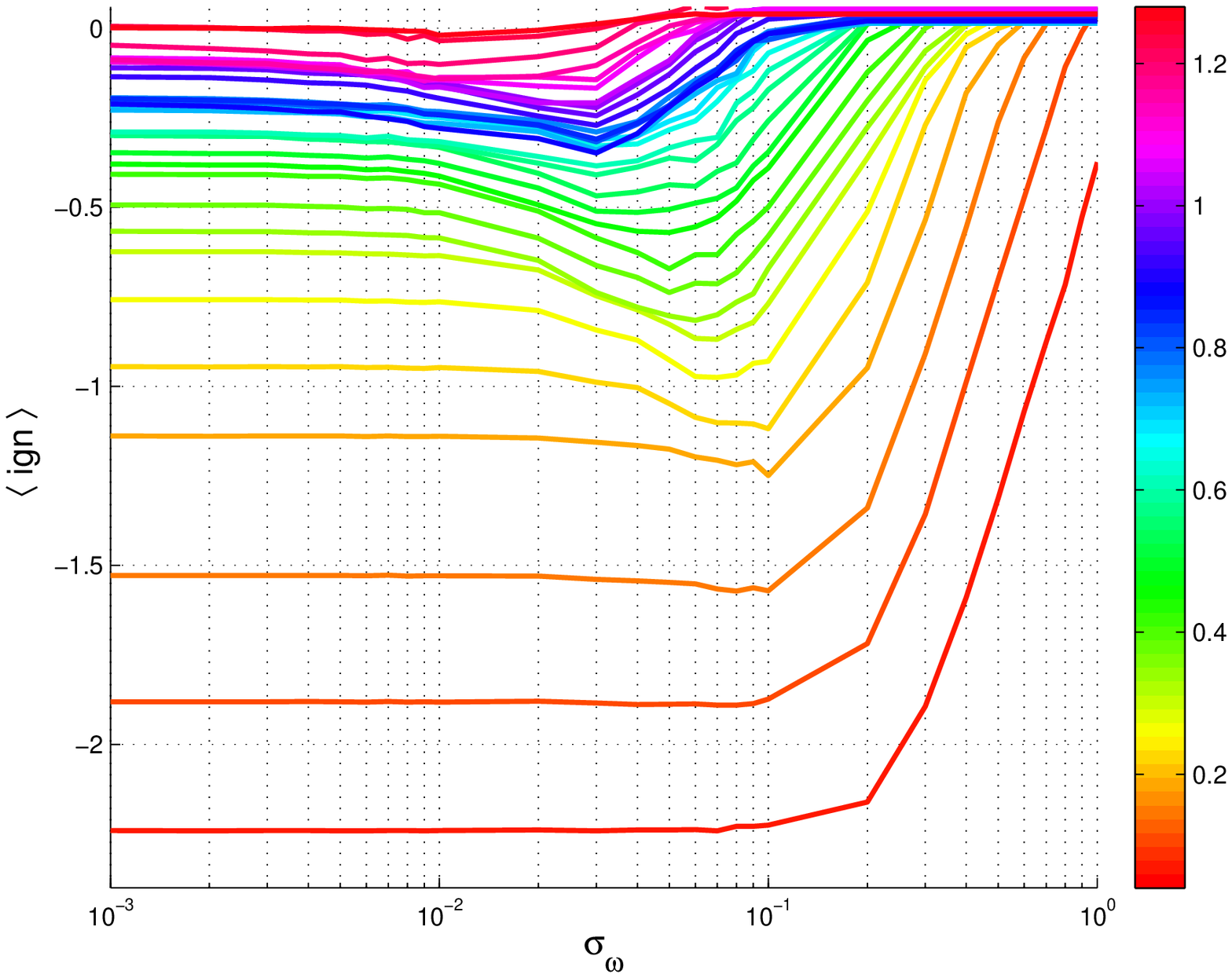}
\includegraphics[width=7.5cm,height=7.5cm]{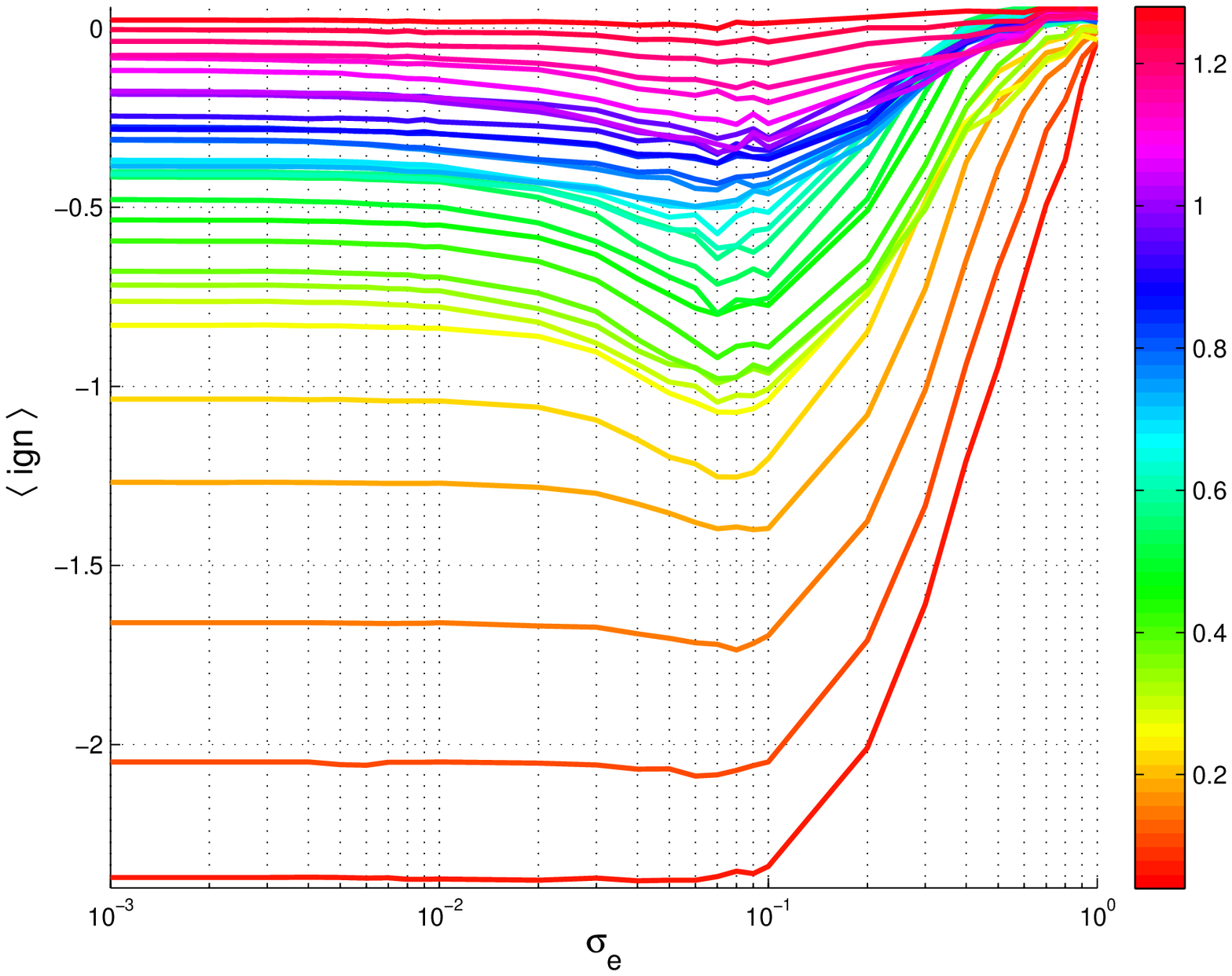}
\caption{\small{Graphs of the average Ignorance score versus dynamical noise spread (left) and initial distribution spread (right)  with observational error of standard deviation $10^{-1}$ on Moore-Spiegel data with an imperfect model. 512 initial conditions with a time step of 32 between them were used. Each initial ensemble containing 33 members was iterated forward up to 32 time steps. The multiple lines correspond to different lead times according to the colorbars.}}
\label{jan3}
\end{figure}

We now turn to the scenario in which there is observational error. If the spread of observational error exceeds that required to yield best forecasting performance in the noise-free case, we will then say observational error dominates model error. We now consider Moore-Spiegel data with observational error of spread $\delta=10^{-1}$. In this case, observational error dominates model error because $\delta>\sigma_{\omega}^*$, where $\sigma_{\omega}^*$ is the optimal spread of dynamical noise in the noise-free scenario. Again we consider forecasts obtained using the model with dynamical noise when the initial conditions are not perturbed and the model without dynamical noise, but with the initial conditions perturbed. The graphs of the average Ignorance score versus the spread of dynamical noise, $\sigma_{\omega}$, and the initial distribution spread, $\sigma_e$, for various lead times are shown in figure~\ref{jan3}.
Firstly, we notice that the model with dynamical noise generally performs worse than the one where only the initial conditions are perturbed. This is especially evident at lower lead times. Moreover, the spread of the dynamical noise that yields the best performance varies with lead time. On the other hand, for the model with no dynamical noise, there is one initial distribution spread that yields the best performance over the range of lead times and this spread turns out to be $\sigma_e\approx\delta$. These points suggest that when observational error dominates model error, perturbing the initial conditions is preferable to using the model with dynamical noise. This contrasts the previous case in which there was no observational error. 

On the right panel of figure~\ref{jan3}, we notice that the low lead time graphs begin to rise at $\sigma_e\approx10^{-1}$. At this value of $\sigma_e$, graphs for the higher lead times dip to reach their minima. This is very much reminiscent of the perfect model scenario and suggests a way of using nonlinear prediction to detect the spread of observational error. Further more, in the face of observational error, model error results in earlier loss of predictability than in the perfect model scenario. In particular, predictability is lost at the highest lead time considered (i.e. $32\times\delta t$) in the imperfect model scenario, which is not the case in the perfect model scenario. The lead time when this happens is about 30 time steps and corresponds to the time for the loss of observational information.

\begin{figure}[!t]
\hspace{-0.6cm}
\includegraphics[width=7.5cm,height=7.5cm]{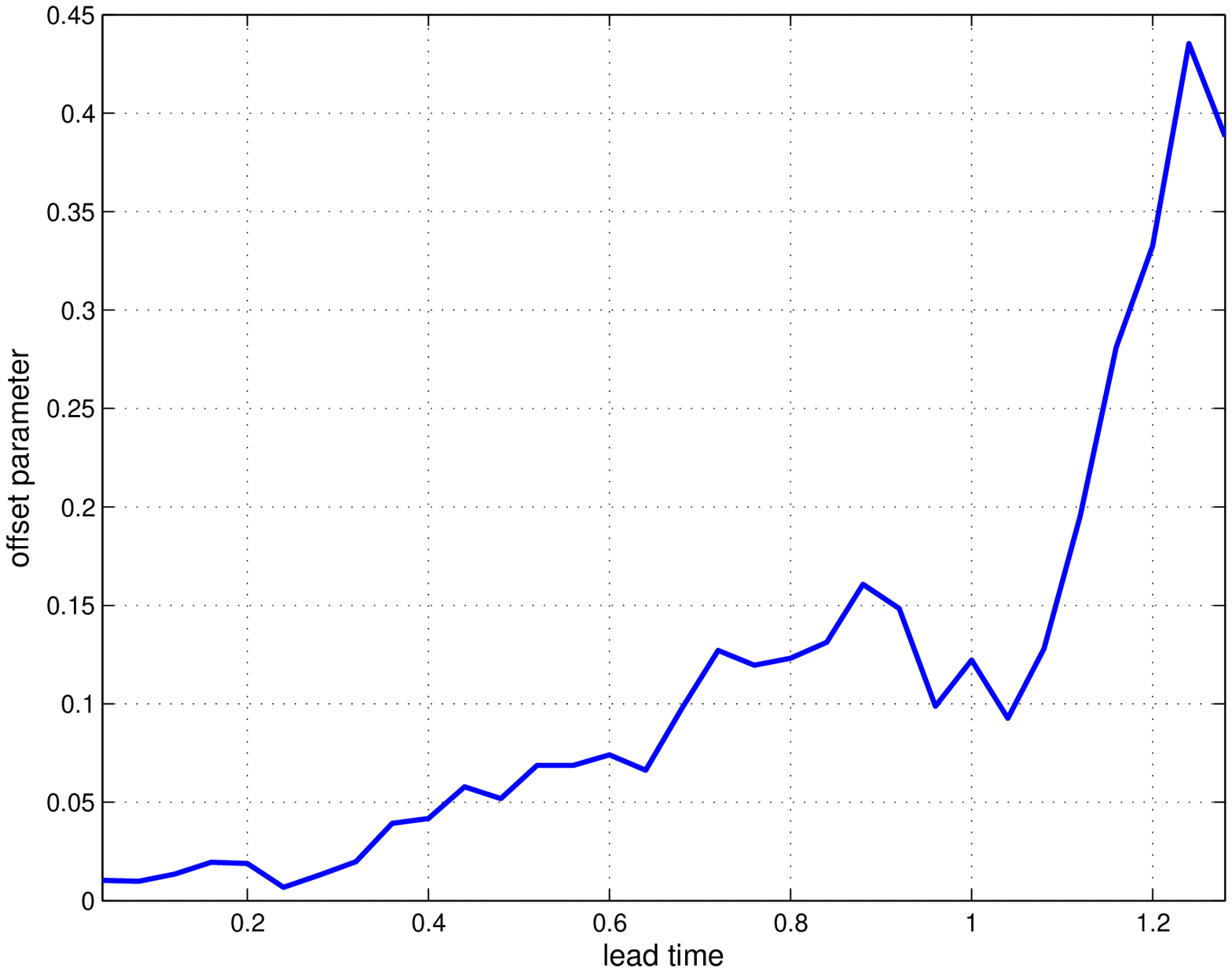}
\includegraphics[width=7.5cm,height=7.5cm]{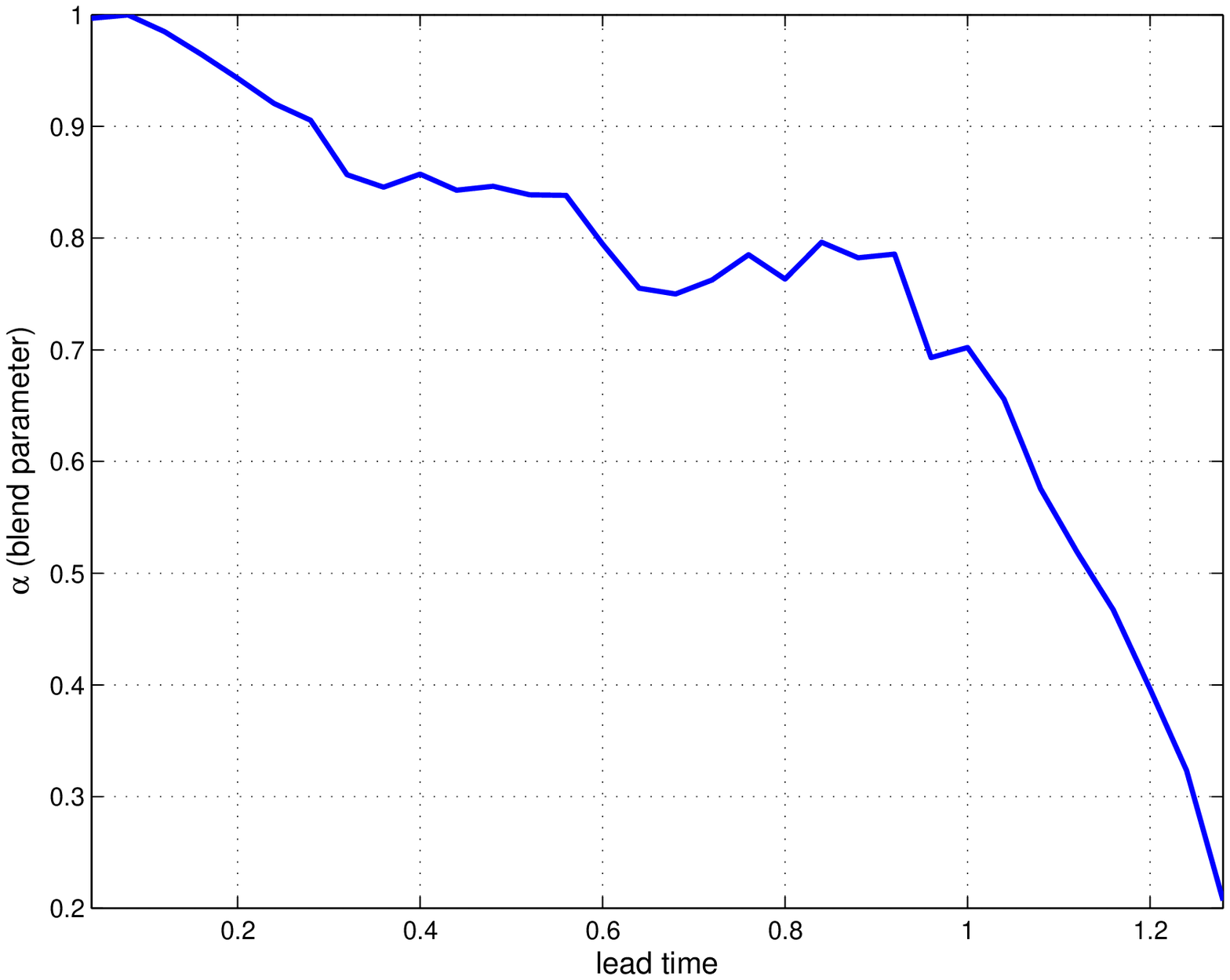}
\caption{\small{Graphs of offset and blend parameters in the density forecasts corresponding initial distribution spread that minimised the average Ignorance score when the observational error was Gaussian with standard deviation, $\delta=10^{-1}$. The initial distribution spread was $\sigma_e=9\times10^{-3}$.}}
\label{imp:fig2}
\end{figure}

Note that when selecting the initial distribution spread with a model that has no dynamical noise, model error is accounted for in the way the density forecasts are formed. In particular, graphs of the offset and blend parameters, corresponding to $\sigma_e\approx\delta$, versus lead time are shown in figure~\ref{imp:fig2}. These parameters are for forming density forecasts according to equation~(\ref{dfe:eq2}) and the graphs correspond to $\sigma_e\approx 10^{-1}$ on the right panel of figure~\ref{jan3}. Tuning these parameters ensures that we do not unduly select a large initial distribution spread because of model error. Notice that the offset parameter is positive and generally increasing with lead time, indicating that the model is biased in one direction at all the lead times. The graph of the blend parameter against lead time, shown on the right panel, provides complementary information. According to the graph, the blend parameter is generally decreasing with lead time. This reflects the degrading value of the forecasting model with lead time. On a positive note, these graphs reflect the amount of correction that was necessary to enhance the quality of the model. 

\begin{figure}[!t]
\centering
\includegraphics[width=0.8\textwidth]{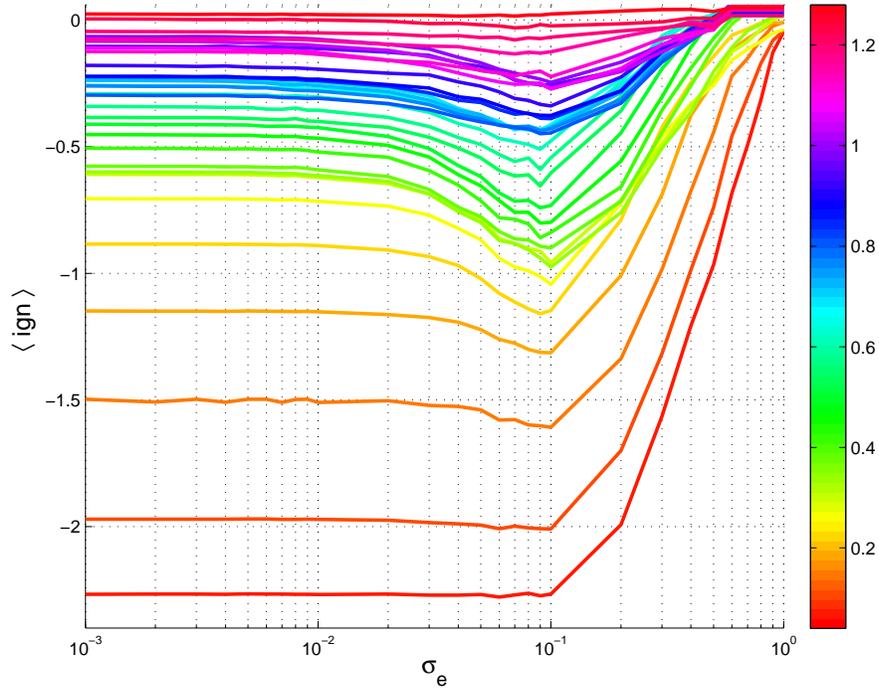}
\caption{\small{Graphs of the average Ignorance score versus logarithmically varying initial distribution spread, $\sigma_e$, of ensemble perturbations with uniformly distributed observational error of standard deviation $\delta=10^{-1}$ on Moore-Spiegel data with an imperfect model. 512 initial conditions with 32 time steps between them were used. From each initial condition, 33 initial ensembles were generated and iterated forward up to 32 time steps. The multiple lines correspond to different lead times according to the colour bar.}}
\label{uni}
\end{figure}

 Suppose now that the observational error of the underlying system is not Gaussian but uniformly distributed with standard deviation $\delta$. We consider this case when the forecasting model has no dynamical noise and assume the noise distribution to be $U[-b,b]$ with $\delta^2=b^2/3$. We have plotted graphs of the average Ignorance score versus initial distribution spread in figure~\ref{uni}
 with $\delta=10^{-1}$. Again we see graphs dipping at $\sigma_e\approx\delta$. While this resembles the case when the observational error and the initial distribution were both Gaussian, it yields higher values of the average Ignorance score. 

The foregoing discussions can be summarised as follows: When observational error dominates model error, the imperfect model scenario with no dynamical noise yields graphs of the average Ignorance score that are similar to those obtained in the perfect model scenario. In the absence of observational error, the imperfect model scenario yields graphs of the average Ignorance score that differ from the perfect model scenario. Graphs of the average Ignorance score versus initial distribution spread (or dynamical noise spread) do not show a linear rise if we are either in the imperfect model scenario or there is observational error. This furnishes us with a simple, heuristic test of whether or not we are in the perfect model scenario without observational error. If the observational error dominates model error, we can detect the spread of the observational error. 
\subsubsection{The Circuit}
In this section we consider an electronic circuit constructed to mimic the Moore-Spiegel equations. Voltages corresponding to the variables $x$, $y$ and $z$ were measured at three different points on the circuit using an instrument called Microlink. In order to minimise ambient temperature effects, the circuit was encased in a metallic box which was then placed in a bigger insulated box prior to data collection. The data was sampled at a frequency of 10$kH$. In this subsection, we considered modelling the circuit only from the voltage signal corresponding to the $z$ variable. Thus we constructed radial basis function models on delay reconstructions as discussed in section~\ref{sec:rbf}. The data was used as collected without preprocessing.

\begin{figure}[!t]
\hspace{-0.6cm}
\includegraphics[width=7.5cm,height=7.5cm]{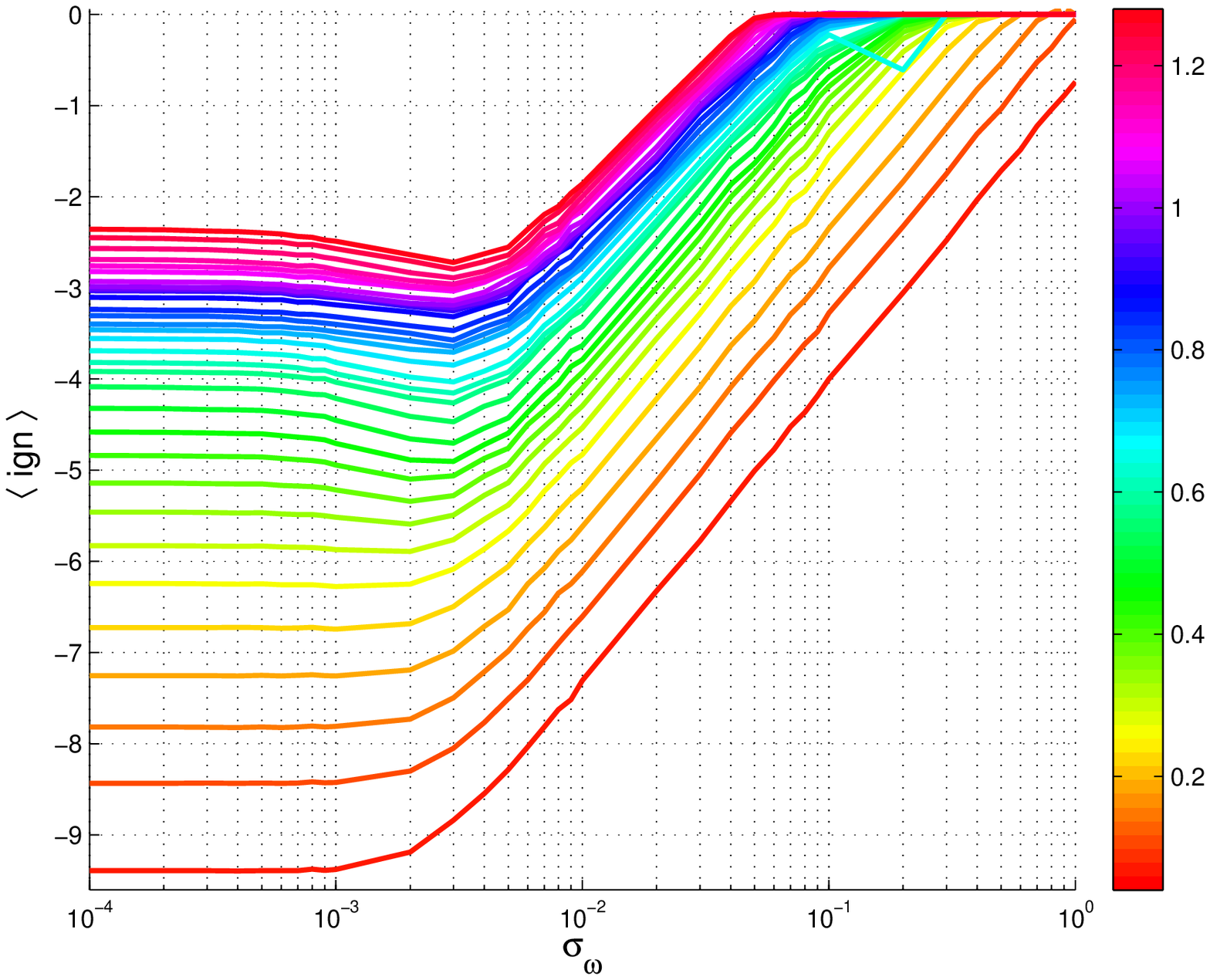}
\includegraphics[width=7.5cm,height=7.5cm]{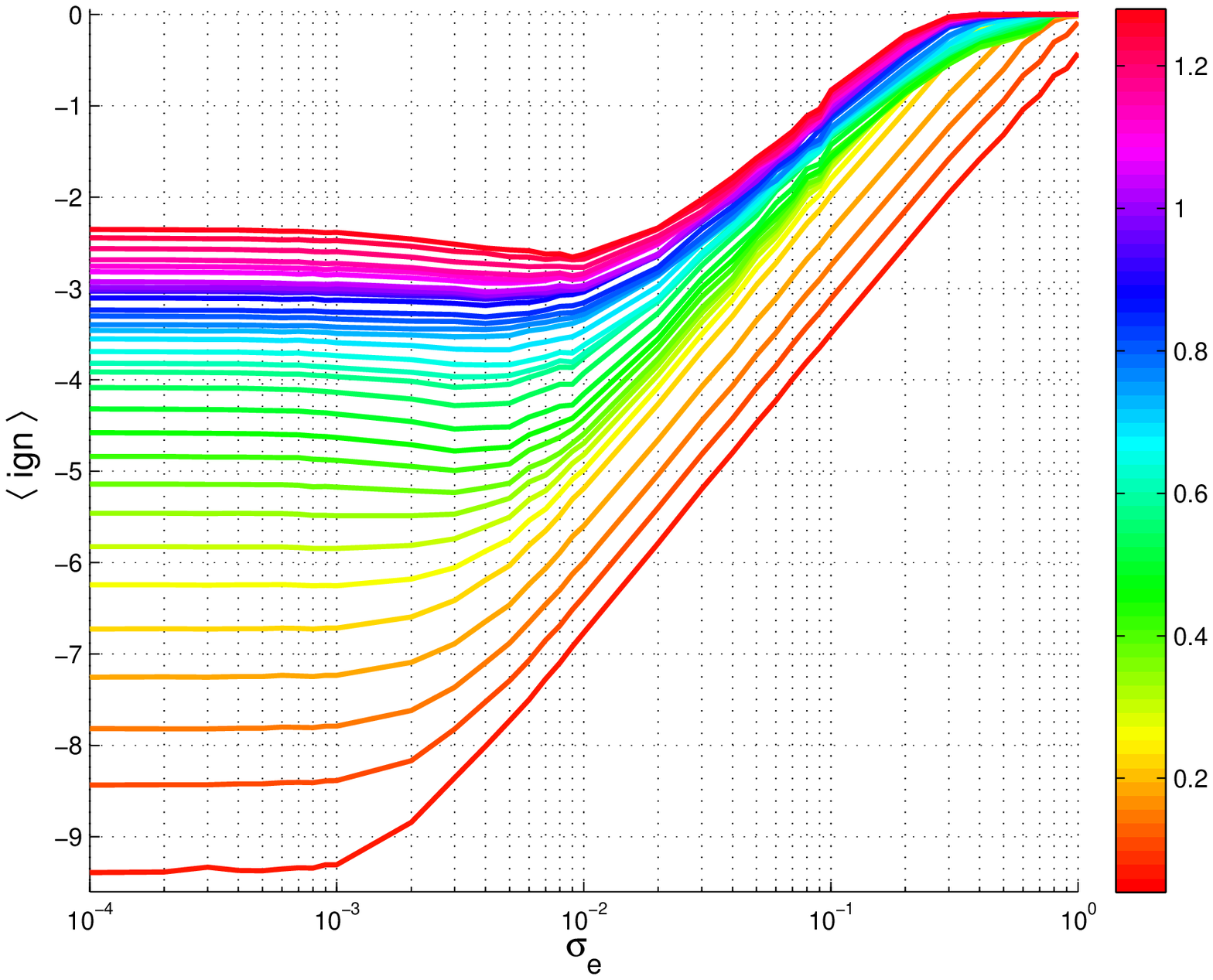}
\caption{\small{Graphs of the average Ignorance score versus dynamical noise spread (left), $\sigma_{\omega}$, and initial distribution spread (right), $\sigma_e$, on circuit data. 512 initial conditions with 32 time steps between them were used. Each initial ensemble containing 32 members was iterated forward up to 33 time steps. The multiple lines correspond to different lead times according to the colour bar.}}
\label{ensc}
\end{figure}

The main question we wish to answer for the circuit is: what distributional spread should we use for a given model? This question is addressed using the average Ignorance score as we have explained in the preceding subsections. We consider cases of the radial basis function model both with and without dynamical noise. Graphs of the average Ignorance score versus dynamical noise spread (left) and initial distribution spread (right) for the circuit are shown in figure~\ref{ensc}. In both cases, the low lead time graphs begin to rise at $\sigma_{\omega}\approx 2\times10^{-3}$ and $\sigma_e\approx10^{-3}$ respectively, which is quite small. Further more, the higher lead time graphs all appear to dip at $\sigma_{\omega}\approx2\times10^{-3}$, for the stochastic model. This suggests that the spread of the underlying observational error is very low. That is, model error dominates observational error and, as would be expected in such a situation, the stochastic model outperforms the deterministic one at the optimal spread. The graphs look much like those obtained with Moore-Spiegel data without observational error, but with an imperfect model~(see figure~\ref{jan3:e0}). These results affirm the accuracy of the data acquisition equipment used to collect the data.

As the spread increases, the graphs flatten out due to performance being poor and climatology taking over. In fact, we estimated the Lyapunov exponent to be $\lambda=0.0281\pm0.0009$ per time step. This implies that the doubling time of initial errors (or uncertainties) is 24 time steps. These observations suggest that the circuit is predictable within the time scale considered. In this case, the forecasts are useful as long as the average Ignorance score is less than $2.15$.
\subsection{Theoretical Considerations}
To explain the previous observations, we consider two pdfs, one of the perfect forecasts and one of the imperfect forecasts: $p_t(x;\sigma_p,\mu_p)$ and $f_t(x;\sigma_f,\mu_f)$. Here $\sigma_p$ (or $\sigma_f$) and $\mu_p$ (or $\mu_f$) are the standard deviation and mean respectively of $p_t$ (or $f_t$), assuming that
\begin{equation*}
\sigma_p(t)=h_p(\sigma_e,t)\quad\mbox{and}\quad\sigma_f(t)=h_f(\sigma_e,t), 
\end{equation*}
where $\sigma_e$ is the initial distribution spread. We will assume that $h_p$ and $h_f$ are increasing functions of $\sigma_e$. Suppose our forecast, $f_t$, is Gaussian, so that
$$f_t(x;\sigma_f,\mu_f)=\frac{1}{\sigma_f\sqrt{2\pi}}e^{-(x-\mu_f)^2/2\sigma_f^2}.$$
Then the expected skill of $f_t$ is
\begin{eqnarray}
\nonumber
\mathbb{E}[\mbox{ign}(f_t,X)]&=&-\int_{-\infty}^{\infty}p_t(x;\sigma_p^2,\mu_p)\log f_t(x;\sigma_f^2,\mu_f)\ud x\\
        &=&\frac{1}{2}\log(2\pi\sigma_f^2)+\frac{\sigma_p^2}{2\sigma_f^2}+\frac{1}{2\sigma_f^2}(\mu_p-\mu_f)^2.
\label{pert:e1}
\end{eqnarray}
If $\sigma_p=\sigma_f$ then~(\ref{pert:e1}) reduces to
\begin{equation}
\mathbb{E}[\mbox{ign}(f_t,X)]=\frac{1}{2}\log(2\pi e\sigma_f^2)+\frac{1}{2\sigma_f^2}(\mu_p-\mu_f)^2.
\label{pert:e2}
\end{equation}
If, in addition, $\mu_p=\mu_f$, then~(\ref{pert:e2}) reduces to
\begin{equation*}
\mathbb{E}[\mbox{ign}(f_t,X)]=\frac{1}{2}\log(2\pi e\sigma_f^2),
\end{equation*} 
which is a monotonically increasing function of $\sigma_f$. This may explain why straight line graphs were obtained in the noise-free perfect model scenario. They arise when the perfect and the imperfect forecasts have equal means and variances.

When there is either model error or observational error, then $\mu_p\neq\mu_f$. The expected skill is then minimised by $\sigma_f^2=(\mu_p-\mu_f)^2$,
with the global minimum given by
\begin{equation*}
\min_{\sigma_f>0}\mathbb{E}[\mbox{ign}(f_t,X)]=\frac{1}{2}\log\left[2\pi e^2(\mu_p-\mu_f)^2\right].
\end{equation*}
In particular, 
\begin{equation*}
\min_{\sigma_e>0}\mathbb{E}[\mbox{ign}(f_0,X)]=\frac{1}{2}\log\left[2\pi e^2\xi_0^2\right],
\end{equation*}
where $\xi_0=\mu_p(0)-\mu_f(0)$ is the observational error, with $\mathbb{E}[\xi_0]=0$ and $\mathbb{E}[\xi_0^2]=\delta^2$. For a more general case, at lead time $t$, we let $\xi_t=\mu_p(t)-\mu_f(t)$. If we let $\mathbb{E}[\xi_t^2]=\sigma_{\xi_t}^2$, then taking the expectation of~(\ref{pert:e2}) with respect to the random variable $\xi_t$ yields
\begin{equation}
\mathbb{E}\{\mathbb{E}[\mbox{ign}(f_t,X)]\}=\frac{1}{2}\log(2\pi e\sigma_f^2)+\frac{1}{2\sigma_f^2}\sigma_{\xi_t}^2,
\label{pert:e6}
\end{equation}
which is minimised by $\sigma_f(t)=\sigma_{\xi_t}$. The corresponding global minimum is
\begin{equation*}
\min_{\sigma_f>0}\mathbb{E}\{\mathbb{E}[\mbox{ign}(f_t,X)]\}=\frac{1}{2}\log(2\pi e^2\sigma_{\xi_t}^2),
\end{equation*}
assuming that $\mathbb{E}[\xi_t]=0$. Since we are considering a chaotic system, we may assume that $\xi_t=\xi_0e^{\lambda t}$, which implies that $\sigma_{\xi_t}=\delta e^{\lambda t}$. It follows that $\sigma_f(t)=\sigma_{\xi_t}$ when $\sigma_f(0)=\delta$. Hence the minimum in~(\ref{pert:e6}) is achieved when the initial distribution spread equals the observational error spread.

It is important to note that an imperfect forecast may be due to an imperfect initial distribution, even if the model is perfect, or it may just be due to an imperfect model regardless of the initial distribution. In the foregoing discussion, we assumed the forecast density to be Gaussian. Whereas this may not be the case for nonlinear systems, it provides analytical and computational tractability. We can consider each Gaussian distribution to be an approximation of the forecast distribution under the model dynamics.
\section{Discussion}
\label{sec:disc}
The computational results presented in this paper demonstrate a way to select the spread of the distribution from which to sample an initial ensemble of points. The goal was to obtain an initial ensemble that would minimise uncertainty in the forecast distributions. The forecasting model need not be perfect for the method to be applied. Information theoretic approaches were used to obtain  the computational results and justify them. The methodology is a departure from traditional data assimilation and ensemble forecasting techniques in a number of ways. We recognise that the ultimate goal of any method that estimates an initial distribution is to obtain more accurate forecasts.

Data assimilation techniques either focus on estimating the true state of the system or finding a set of such estimates. To this end, a model trajectory may be sought that is consistent with observations~\cite{kev-01,bur-98}. It is believed that forecasts made from an ensemble that lies along such trajectories would provide good forecasts. An ensemble of trajectories is obtained by making perturbations of some initial observation. When there is model error, there is no model trajectory that is consistent with observations. Therefore, Judd and Smith~\cite{kev-04} talk of pseudo-orbits instead. Notwithstanding these difficulties, the method presented  here could be used to determine the spread of this distribution, regardless of the data assimilation technique. For a given structure of the correlation matrix, we would seek the scalar multiple that yields the most informative forecast distributions.

Other techniques for producing the initial ensemble aim at selectively sampling those points that are dynamically the most relevant. In particular, the ECMWF ensemble prediction system seeks perturbations of the initial state based on the leading singular vectors of the linear propagator~\cite{leu-08}. This approach can lead to over-confidence when there is model error. One falls into the trap of confusing the dynamics of the model with those of the underlying system as highlighted in~\cite{len-chaos}. To safeguard this problem, our methodology may be used to select the initial distribution spread. 

The results also suggest that the method may be useful in nonlinear noise reduction, where the quality of the model would have to be very good, at least in the sense of forecasting. However, the primary value of the method is to find the spread of the initial distribution. It is also interesting that even when there is no observational error, sampling the initial distribution could still help mitigate model inadequacy. However, there is a problem that the optimal spread varies with lead time. In such a case, while the stochastic approach advocated for by~\cite{jud-07} provides a superior alternative, the method presented here provides guidance on how to choose the spread of the stochastic perturbations.

Finally, possible areas of application go beyond Meteorology and the Geosciences. For instance, evidence of nonlinear dynamics has already been reported in Economics and Finance~\cite{tev-06}. The Bank of England Quarterly model, which is nonlinear~\cite{harr-05}, is another example. In some cases the dynamics are fairly low dimensional~(e.g. \cite{lin-93}), thus reducing the computational costs that may arise from generating an initial ensemble. We envision the method being of great value in these disciplines to tackle density forecasting.
\section{Conclusions}
\label{sec:con}
This paper argued for combining the task of choosing the initial ensemble with density forecasting. The point is that, when faced with model error, a knowledge of the true state of the system is irrelevant because it cannot provide one with a perfect forecast. Moreover, using the true state with an imperfect model can provide forecasts that are further from the truth than forecasts obtained with imperfect initial states. Therefore, it has been argued that the task of the forecaster should be to choose initial distributions that yield the most informative forecast distributions. Whereas this approach may be incorporated into traditional ensemble forecasting techniques, it can also stand independently as a forecasting method.

To recap, it was demonstrated that the logarithmic scoring rule can be used to estimate an optimal initial distribution spread for a given system and model. At the optimal spread, higher lead time graphs of the logarithmic scoring rule versus initial distribution spread tend to dip. The distribution of the underlying observational uncertainty or model error seems not to play a crucial role. It turns out that we can also diagnose the fictitious case of a perfect model with perfect initial states. A theoretical explanation for the empirical observations regarding the dipping of the graphs has been presented.
\section*{\bf\Large Acknowledgements}
The authors would like to thank Leonard A. Smith, Devin Kilminster and members of the Applied Dynamical Systems Group at Oxford for fruitful discussions and useful insights. Comments from four anonymous reviewers have helped improve the manuscript. This work was supported by the LSE's Higher Education Infrastructure fund and the RCUK Digital Economy Programme fund at the University of Reading via EPSRC grant EP/G065802/1 The Horizon Digital Economy Hub.
\bibliographystyle{elsarticle-num}
\bibliography{refs}
\appendix
\section{Invariant Density}
\label{app:inv}
Associated with the attractor $A$ is some {\it invariant measure}~\cite{er-85}, $\varrho$, such~that
\begin{equation}
\varrho[\boldsymbol{\varphi}_{-t}(E)]=\varrho(E),
\label{s1:eq2}
\end{equation}
where $E\subset\mathbb{R}^m$ is a measurable set and $\boldsymbol{\varphi}_{-t}(E)$ is the set obtained by evolving each point in $E$ backwards in time. A probability measure on $E$ may be defined as~\cite{er-85}
\begin{equation}
\varrho(E)=\lim_{T\rightarrow\infty}\frac{1}{T}\int_0^T\boldsymbol{1}_E(\boldsymbol{\varphi}_t(\boldsymbol{x}_0))\ud t,
\label{s1:eq3}
\end{equation}
where $\boldsymbol{1}_E$ is an indicator function~\footnote{An indicator is defined by $$\boldsymbol{1}_E(\boldsymbol{x})=\left\{\begin{array}{ll}
1 & \mbox{if}\quad \boldsymbol{x}\in E,\\
0 & \mbox{if}\quad\boldsymbol{x}\not\in E.
\end{array}\right.$$}. 
Provided the attractor $A$ is {\it ergodic}~\footnote{In an ergodic attractor, state space averages are equal to time averages~\cite{er-85}.}, 
\begin{equation}
\varrho(E)=\int_E\varrho(\ud\boldsymbol{x}).
\label{s1:eq4}
\end{equation}
Associated with $\varrho$ is some probability density function, $\rho$, such that~(\ref{s1:eq4}) may be rewritten as
\begin{equation}
\varrho(E)=\int_E\rho(\boldsymbol{x})\ud\boldsymbol{x}.
\label{s1:eq5}
\end{equation}
We call $\rho(\boldsymbol{x})$ the {\it invariant density} of the attractor $A$ or the flow $\boldsymbol{\varphi}_t(\boldsymbol{x}_0)$. This invariant density is indeed the climatology~\footnote{Including its marginal densities.} mentioned in the introduction.
\end{document}